\documentclass[fleqn,usenatbib]{mnras}
\usepackage{newtxtext,newtxmath}
\usepackage[T1]{fontenc}
\usepackage{ae,aecompl}
\usepackage{graphicx}    
\title[Thermal Relaxation of DMANS] {Thermal Relaxation of Dark Matter Admixed Neutron Star}
\author[Kumar et al.]{
Ankit Kumar$^{1,2}$
\thanks{ankit.k@iopb.res.in},
H. C. Das$^{1,2}$
S. K. Patra$^{1,2}$,
\\
$^{1}$ Institute of Physics, Sachivalaya Marg, Bhubaneswar-751005, India\\
$^{2}$ Homi Bhabha National Institute, Training School Complex, 
Anushakti Nagar, Mumbai 400094, India}
\begin{document}
\maketitle
\begin{abstract}
Motivated by the various theoretical studies regarding the efficient capturing of dark matter by neutron stars, we explore the possible indirect effects of captured dark matter on the cooling mechanism of a neutron star. The equation of states for different configurations of dark matter admixed star at finite temperature is obtained using the relativistic mean-field formalism with the IOPB-I parameter set. We show that the variation in the dark matter momentum vastly modifies the neutrino emissivity through specific neutrino generating processes of the star. The specific heat and the thermal conductivity of a dark matter admixed star have also been investigated to explore the propagation of cooling waves in the interior of the star. The dependence of theoretical surface temperature cooling curves on the equation of state and chemical composition of the stellar matter has also been discussed along with the observational data of thermal radiation from various sources. We observed that the dark matter admixed canonical stars with $k_{f}^{\rm DM} > 0.04$ comply with the fast cooling scenario. Further, the metric for internal thermal relaxation epoch has also been calculated with different dark matter momentum and we deduced that increment of dark matter segment amplify the cooling and internal relaxation rates of the star.
\end{abstract}
\section{Introduction}\label{intro}
Neutron stars, undoubtedly one of the most astounding objects in the universe, hold infinitude of possibilities in terms of density profile, structure, symmetry, extreme surface temperature, magnetic field and composition, which have been barely investigated using our limited acquirable knowledge. The existence of parallel correspondence between its rearmost stellar and nuclear properties makes it even more enigmatic and interesting. The physical aspects of a neutron star can be explained in a complete and comprehensive manner only if one can connect the drafts of various unassociated subdivisions of physics like nuclear physics, neutrino physics, general relativity, quantum chromodynamics, plasma physics and superfluid dynamics in a synergical manner. The interpretation of pulsars as rotating stars during the numerous observational events established the neutron stars as effectual objects to explore the physics of dense matter systems and laid the foundation for nuclear astrophysics as a distinct field of interest \citep{1968Natur.218..731G}.

The well-established fact about the mysterious and uncanny neutron stars is that the matter in the core of it can reach upto $5-10$ times of the nuclear saturation density. The core is responsible for the major percentage of the neutron star's mass and holds extreme physical phenomena and conditions like neutronization of matter, superconductivity and superfluidity at extremely high temperatures, high magnetic field etc., which are far away from existence in the subastral environment \citep{PhysRevLett.106.081101}. Since, the theories or the model proposed to perceive the physics of matter at such high densities can not be tested in terrestrial laboratories, therefore, the selection of an accurate equation of state (EOS) to describe the structure and composition of neutron star cores is a huge challenge for the physicists in the last few decades. A number of EOSs obtained using different theoretical approaches have been used in the past few years to unearth the physics of neutron stars and to satisfy the constraints put by several observational events on various properties of the neutron star \citep{RevModPhys.81.1773, refId0, doi:10.1146/annurev-nucl-102711-095018, Dutra_2012, PhysRevC.91.014002, PhysRevC.94.064326, Greif_2020}. An elaborated review with detailed discussions on various methods such as meson-exchange and potential models; chiral effective field theory and ab initio methods; Skyrme and Gogny interactions etc. to obtain the equation of state for supernovae and compact stars is done by Oertel et. al. \citep{RevModPhys.89.015007}. One of the most successful and prominent theoretical perspectives to obtain the EOS for the investigation of various properties of a dense star is relativistic mean-field (RMF) formalism. The entire spectrum of the EOS from the stiffest to the softest region can be obtained with the help of existing RMF parameter sets in the literature, which corresponds to a different order of coupling factors between mesons and nucleons and has its own merits in terms of satisfying the various astrophysical observational constraints  \citep{WALECKA1974491, PhysRevC.55.540, PhysRevC.65.035802, PhysRevC.93.025806, KUMAR2017197, PhysRevC.100.025805}.

The composition of the neutron star cores has also been a secret and topic of debate in the last few years. Although, we can infer a broad idea about the neutron star interiors using the astrophysical observations, but it is far from reach to find out the exact creation of the dense star on the basis of observational data. There are a lot of hypothesis available in the literature regarding the availability of free quarks, hyperons, delta isobars, or the phenomenon of kaon condensation in the interior shell of a neutron star \citep{1985ApJ...293..470G, KAPLAN1988273, PhysRevLett.34.1353, PhysRevC.36.1019, PhysRevC.56.1570}. The presence of exotic particles inside the neutron star happens to soften the EOS which in turn will affect the observables, e.g., reduce the mass and tidal deformability of the star \citep{PhysRevLett.67.2414, PhysRevC.85.065802, 10.1093/mnras/stab2387}. Apart from the standard exotic particles, theoretical studies also suggested the presence of dark matter particles inside the neutron star. The ultra-dense profile of the star and asymmetry between particles and anti-particles of dark matter branch is supposed to be responsible for the gravitational trapping of dark matter without self-annihilation in the core of a compact star \citep{PhysRevD.40.3221, PhysRevD.77.023006,2012arXiv1204.2564F}. The accumulated dark matter inside a dense star with either being bosonic or fermionic in nature and with different hypothesized dark matter candidates from the theoretical and experimental point of view such as neutralinos, axions, technibaryons have been explored in various theoretical studies \citep{Jungman_1996, Bertone_2005, Leanne_2009, PhysRevD.82.063531, 2012arXiv1204.2564F, PhysRevC.89.025803}. Although, the nature, direct interaction or coupling of dark matter with the other standard model particles is quite weak and uncertain, but theoretical efforts have been made to explore the effects of dark matter interaction on the properties of astral bodies \citep{LI201270, Li_2012, PhysRevD.93.083009}. The peak in the power spectral density of the gravitational wave observations of the GW170817 event which involves a neutron star-neutron star merger and coalescence event involving a black hole and compact object (GW190814), also offers a new approach for the search and effects of DM on the components of universe \citep{PhysRevLett.119.161101, ELLIS2018607, Abott_2020}.

Another important aspect that can help a lot in understanding the physical processes taking place inside a neutron star is the elaborated study of its cooling dynamics. The first theoretical attempt to study the cooling mechanism of a neutron star was done in the 1960s in order to establish them as the newly discovered X-ray source detected by rocket and balloon experiment \citep{PhysRevLett.9.439, PhysRevLett.12.413, doi:10.1126/science.146.3646.912}. Later, the theoretical cooling simulations extended in a more comprehensive manner within the context of the theory of general relativity for a hydrostatic spherically symmetric neutron star, which ultimately accelerated the importance of physical factors of neutron stars in X-ray astronomy \citep{1980ApJ...239..671G, 1986ApJ...305L..19N, 1981ApJ...244L..13V}. The study of the thermal evolution of a neutron star has different dimensions engaged with mass accretion, composition, magnetic field, superfluidity, superconductivity etc. Theoretical studies suggest that the temperature of a newly born proto-neutron star is around $10^{10} \,K$ and the initial rate of heat loss is so intense that the temperature drips down to $10^{9}-10^{8} \,K$ within a few seconds. The excessive  neutrino flux is much more dominant than surface photon radiation and mainly responsible for the initial stage rapid cooling of a neutron star \citep{1986ApJ...307..178B, Pons_1999}. As a matter of fact, the neutrino flux is so enormous that even the Pauli effects on the fermionic character of the neutrinos control the the dynamics of a supernova explosion considerably \citep{1975ApJ...195L..19B}. The superfluid effects can also influence neutrino emissivity and the phenomenon of heat transport within the shell of a neutron star. Superfluidity leads to the Cooper pairing of nucleons which can enhance the neutrino emissivity by generating more neutrinos and hence can influence the thermal evolution of the star \citep{Yakovlev_1999}. The composition of a dense star controls the production of neutrinos via different processes inside the star, which indirectly influence its cooling mechanism \citep{10.2307/53909, PAGE2006497}. The high magnetic field of the neutron star is responsible for the anisotropy in the thermal conductivity, which ultimately evolves into the surface temperature gradient. Neutron stars remains apparent up to million years via surface thermal emission and the thermal imaging of steady X-ray spectra from the supernova remnants using the imaging proportional counter of orbital X-ray observatories (such as Einstein Observatory, Chandra X-ray Observatory, XMM-Newton, MAXI, NuSTAR, Astrosat, Max Valier Sat, NICER, eROSITA, Spektr-RG etc.) can provide a lot of useful information about the effective surface temperature of the stars \citep{1979ApJ...234L...1G, BOGUTA1981255}. The temperature-susceptive properties such as electron thermal conductivity, specific heat, crust solidification etc., can also be explored by deciphering the thermal evolution of the star. The outer layers of the neutron star, generally known as the crust, control the surface photon emission and yield a lot of information about the mechanism of thermal relaxation inside the envelope of the star. The simple assumption about the ingredient of the envelope of a neutron star i.e. whether the envelope is made up of light elements or heavy elements, can subvene the surface temperature versus relaxation time trajectory of the star quite considerably \citep{BEZNOGOV20211}.

Theoretical studies suggest that the accretion of weakly interacting massive particles (WIMPs) onto the neutron star can induce heating in the aged stars, which can be explored to put several constraints on the cross-section of the dark matter inaccessible to the experiment \citep{PhysRevD.40.3221}. The accretion of interstellar gas and massive particles could possibly trigger the reheating phenomenon inside the aged ($10^{6}$-$10^{7}$ years old) neutron stars and ultimately lead its fate into a collapse \citep{PhysRevD.77.023006, PhysRevD.97.043006}. Also, the annihilation of WIMP in the core of a neutron star can act as the energy source required for the phase transition to quark matter and considerably affect the dynamics of supernova explosion \citep{Hong, PhysRevD.81.123521, PhysRevD.77.023006}. However, in the present work, we did not consider the heating effect due to WIMPs or the energy bump due to the dark matter annihilation which corresponds to inappreciable cross-section, as done in the above-cited references. In this work, we revisit the dynamics of thermal relaxation for a neutron star in the context of nucleon interaction with dark matter postulator indirectly. We explore the effects of indirect interaction of dark matter on neutrino emissivity, specific heat, thermal conductivity and surface temperature of the neutron star. The paper is commenced with the formalism used to obtain the EOS of neutron stars along with the effect of dark matter interaction. We derive the EOS of a dark matter admixed neutron star using the temperature-dependent RMF formalism with different variations of dark matter momentum. In the later sections, the obtained EOS is used to study the mechanism of thermal relaxation of a neutron star.
\section{Equation of State}\label{EoS}
The RMF theory, introduced in the context of infinite nuclear matter, later proved to be one of the most successful formalism to study the properties of finite nuclei and compact astrophysical objects \citep{Serot1992, GAMBHIR1990132, RING199777, PhysRevC.55.540}. The Lagrangian density introduced initially evolved with the addition of new interactions and multi-mesonic couplings, to satisfy several constraints imposed by the experimental and empirical data. Also, a number of RMF parameter sets are developed by different group of researchers, each one of which has its own importance and demerits in view of the flow data experiment and observational constraints \citep{PhysRevC.55.540, FURNSTAHL1997441, PhysRevC.66.055803, PhysRevLett.95.122501, PhysRevC.74.034323, PhysRevC.76.045801, PhysRevC.82.055803, PhysRevC.90.055203}. We adopted the newly developed RMF parameter set by Kumar {\it et al.} named as "IOPB-I" \citep{PhysRevC.97.045806} and for which the corresponding Lagrangian density representing the hadronic matter for a neutron star subsequent to mean-field approximation can be written as \citep{2005EPJA...25..293L, 2017nuco.confb0811O, KUMAR2021122315}
\begin{eqnarray}\label{lag}
{\cal{L}}_{NM} &=& \sum_{j=p,n}\bar\psi_{j}
\Bigg\{\gamma_{\mu}(i\partial^{\mu}-g_{\omega}\omega^{\mu} -\frac{1}{2}g_{\rho}\vec{\tau}\!\cdot\!\vec{\rho}^{\,\mu})
\nonumber\\
& &
-(M-g_{\sigma}\sigma)\Bigg\}\psi_{j}
-\frac{1}{2}m_{\sigma}^{2}\sigma^2
+\frac{1}{2}\partial^{\mu}\sigma\,\partial_{\mu}\sigma
\nonumber \\
& & 
-g_{\sigma}\frac{m_{\sigma}^2}{M}
\Bigg(\frac{\kappa_3}{3!}
+ \frac{\kappa_4}{4!}\frac{g_{\sigma}}{M}\sigma\Bigg)
   \sigma^3
+\frac{1}{2}m_{\omega}^{2}\omega^{\mu}\omega_{\mu}
\nonumber \\
& & 
-\frac{1}{4}F^{\mu\nu}F_{\mu\nu}
+\frac{\zeta_0}{4!}g_\omega^2
   (\omega^{\mu}\omega_{\mu})^2
+\frac{1}{2}m_{\rho}^{2}\rho^{\mu}\!\cdot\!\rho_{\mu}
\nonumber\\
& &
-\frac{1}{4}\vec R^{\mu\nu}\!\cdot\!\vec R_{\mu\nu}
+\Lambda_{\omega}g_{\omega}^2g_{\rho}^2(\omega^{\mu}\omega_{\mu})(\vec\rho^{\,\mu}\!\cdot\!\vec\rho_{\mu})
\nonumber\\
& &
+\bar\phi_{l}\,(i\gamma_{\mu} \partial^{\mu} - m_e)\phi_{l},
\label{eq1}
\end{eqnarray}
where, $M$ and $\psi$ are the mass and wave function for the  nucleons (protons and neutrons) and the last term of the Lagrangian corresponds to the leptons (electrons and muons). The considered masses of the intermediating $\sigma$, $\omega$ and $\vec{\rho}$ mesons for IOPB-I parameter set are $500.512$, $782.5$, and $763$ MeV and the coupling constants of the corresponding mesons i.e. $g_{\sigma}$, $g_{\omega}$ and $g_{\rho}$ are $-10.3954$, $13.3466$ and $11.1242$ respectively. The self and cross-coupling coefficients i.e. $\kappa_{3}$, $\kappa_{4}$, $\zeta_{0}$ and $\Lambda_{\omega}$ for the IOPB-I parameter set are $1.4957$, $-2.9324$, $3.1039$ and $0.0243$ respectively. The EOS for a neutron star acquired by using the IOPB-I parameter set is consistent in view of the nuclear matter saturation properties and satisfy almost all the bars imposed by the gravitational observational phenomenon (GW170817), which can be verified in the references \citep{PhysRevC.97.045806, PhysRevD.104.123006, 10.1093/mnras/stab2387}. Since, we want to explore the thermal mechanism in this work, so, we infer the EOS representing the neutron star at finite temperature. The total nucleonic ($n$) and scalar ($n_{s}$) density within RMF formalism for a finite temperature neutron star will be represented by,
\begin{eqnarray}
n &=& \sum_{j=p,n} \langle\bar\psi_{j}\gamma_0\psi_{j}\rangle  = n_{p}+n_{n} \nonumber\\
&=& \sum_{j=p,n}\frac{2}{(2\pi)^{3}}\int_{0}^{k_{F_{j}}}d^{3}k\,[f_{j}(\mu_{j}^{\ast},T)-\bar f_{j}(\mu_{j}^{\ast},T)],
\label{eq2}\\
n_s &=& \sum_{j=p,n}\langle\bar\psi_{j}\psi_{j}\rangle = n_{sp}+n_{sn} \nonumber\\
&=& \sum_{j=p,n} \frac{2}{(2\pi)^3}\int_{0}^{k_{F_{j}}} d^{3}k\, \frac{M_{j}^{\ast}} {(k^{2}_{j}+M_{j}^{\ast 2})^{\frac{1}{2}}}
\Big[f_{j}(\mu_{j}^{\ast},T)
\nonumber\\
 &&
+\bar f_{\alpha}(\mu_{\alpha}^{\ast},T)\Big].
\label{eq3}
\end{eqnarray}
$M^{\ast} = M - g_{\sigma} \sigma$, is the effective mass of the nucleons in mesonic field and $f_{j}(\mu_{j}^{\ast},T)$ and $\bar f_{j}(\mu_{j}^{\ast},T)$ represents the Fermi distribution function for the particle and anti-particle at effective chemical potential $\mu_{j}^{\ast}$ and temperature $T$. The temperature dependent mean-field mechanism to study the macroscopical properties of a neutron star at finite temperature is explicitly discussed and analyzed in our recent work \citep{Kumar2020, KUMAR2021122315}.

\begin{figure}
\centering
\includegraphics[width=1\columnwidth]{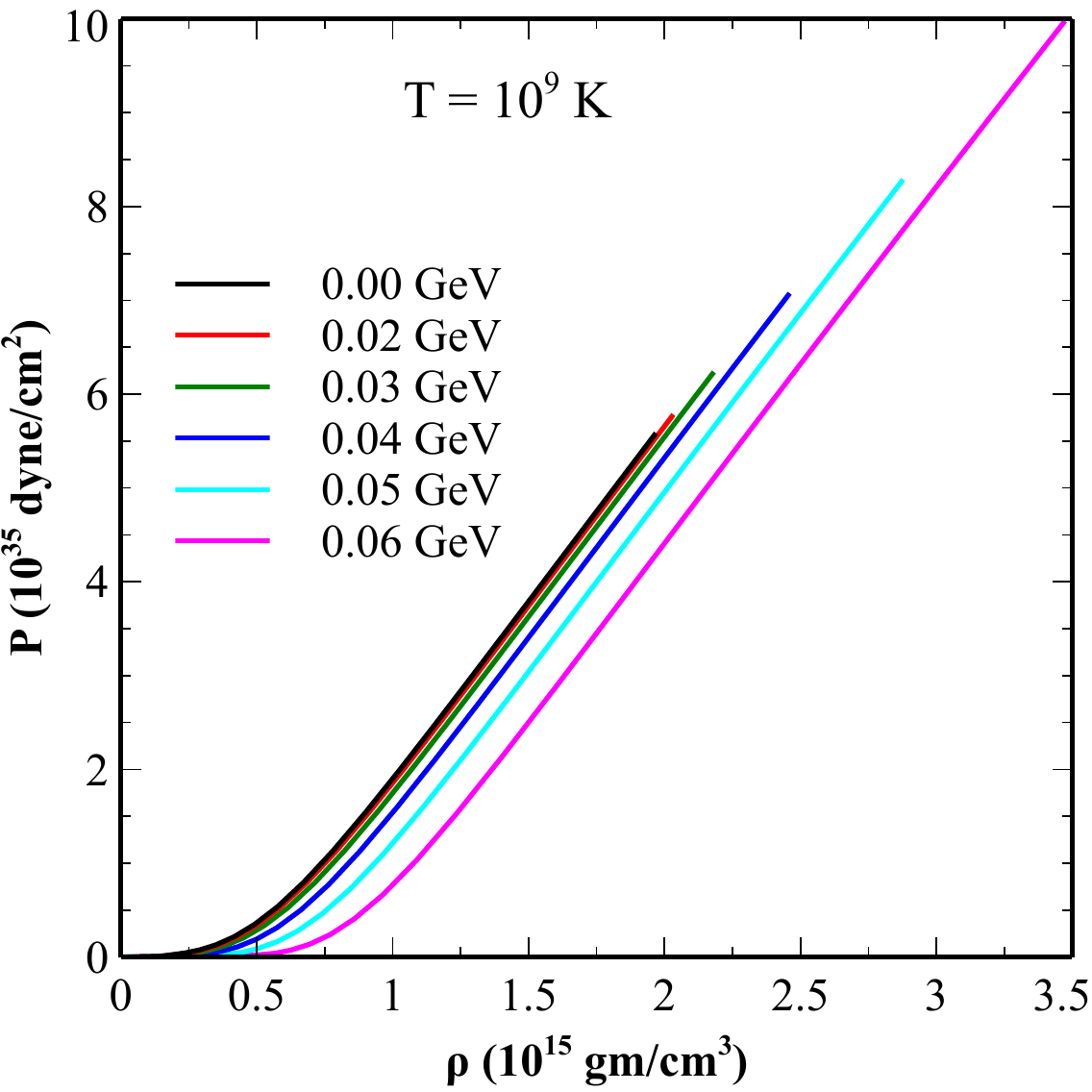}
\caption{Equation of state for a dark matter admixed neutron star at $T = 10^{9}\,K$. The dark matter momentum ($k_{f}^{\rm DM}$) is varied in-between $0.02-0.06$ GeV and the black line curve corresponds to the neutron star EOS without any dark matter component. }
\label{fig1}
\end{figure}

Now, in the present work, along with the temperature we incorporate the effects of trapped fermionic dark matter also on the EOS of the neutron star. The functional Lagrangian density to obtain the EOS for a finite temperature dark matter admixed neutron star can be written as, \citep{PhysRevD.96.083004, PhysRevD.99.043016, 10.1093/mnras/staa1435}, 
\begin{eqnarray}
{\cal{L}} & = & {\cal{L}}_{NM} + \bar \chi \left[ i \gamma^\mu \partial_\mu - M_\chi + y h \right] \chi +  \frac{1}{2}\partial_\mu h \partial^\mu h  \nonumber\\
& &
- \frac{1}{2} M_h^2 h^2 + \frac{f\,M}{v} \bar \psi h \psi , 
\label{eq4}
\end{eqnarray}
where $\chi$ represents the lightest neutralino wave function, which for the present case is assumed as fermionic dark matter candidate \citep{MARTIN_1998}, $h$ stands for the Higgs field and, the last term is the nucleon-Higgs coupling. The detailed expressions for the energy density and pressure using the above defined Lagrangian density are derived in the references \citep{10.1093/mnras/staa1435, lourenco2021dark}. The Yukawa coupling parameter for the nucleon-Higgs interaction ($f\,M/v$) is set to $0.001336$ and since the theoretically defined range for the neutralino-Higgs coupling ($y$) is $0.001-0.1$, so we fix its value to be 0.07 \citep{PhysRevD.88.055025, PhysRevD.88.054507,  PhysRevD.91.075005}. The effective mass of the nucleons in this scenario will be modified as $M^{\ast} = M - g_{\sigma} \sigma -\frac{f\,M}{v} h$. The dark matter density is considered around thousand times smaller than the nuclear saturation density and the composition of dark matter inside the neutron star is controlled through its momentum ($k_{f}^{\rm DM}$), which is varied upto $0.06$ GeV and $k_{f}^{\rm DM}=0.00$ GeV reveals the equation of state of $\beta$ equilibrated matter without any dark matter component \citep{PhysRevD.96.083004, PhysRevD.99.043016}. We stabilize the star's central temperature at $10^{9}\,K$ and also preserve the beta equilibrium and charge neutrality to get the EOS and fractional abundance of particles for a dark matter admixed star. Also, the crustal segment of the star plays an important role in the simulation dynamics of thermal evolution, so we use the stellar crust EOS data derived by Chamel and Haensel \citep{Chamel2008}. The combined EOS for the crust and core of the dark matter admixed neutron star is depicted in Fig. \ref{fig1}. 
We publicized that the increase in the dark matter momentum at zero temperature shifts the EOS towards the softer region which indirectly results in the decreased mass of the star. The interaction between the nucleons and the dark matter particle via Higgs and the inclusion of more degrees of freedom is accountable to make the EOS softer. The system would like to minimize its energy through low energy states of the dark matter, so, the nucleons are replaced by the dark matter candidates depending on the system's density due to which a fraction of the gravitational mass is converted to kinetic energy \citep{10.1093/mnras/staa1435, PhysRevD.104.123006}. A similar pattern for the EOS is observed at the finite temperature (Fig. \ref{fig1}). The maximum mass and the corresponding radius of the neutron star for the EOS at $T = 10^{9}\,K$ using IOPB-I parameter set with different dark matter momentum is calculated using Tolman-Oppenheimer-Volkoff equations \citep{PhysRev.55.364, PhysRev.55.374}. The obtained canonical star radius ($R_{1.4}$) along with the maximum mass and corresponding radius ($M, R$) for the neutron star with $k_{f}^{\rm DM}=0.00 - 0.06$ GeV are stated in table \ref{table1}. 
\begin{table}
\caption{The maximum mass ($M$), radius for maximum mass star ($R_{M}$) and canonical star radius ($R_{1.4}$) for different values of dark matter Fermi momentum ($k_{f}^{\rm DM}$) using IOPB-I parameter set at $10^{9}\,K$.}
\renewcommand{\tabcolsep}{0.5cm}
\renewcommand{\arraystretch}{0.8}
\label{table1}
\begin{tabular}{|c|c|c|c|}
\hline
{$k_{f}^{\rm DM}$} & {Max. Mass ($M$)} & {$R_{M}$} & {$R_{1.4}$} \\ \\ 
(GeV) & ($M_{\odot}$) & (km)& (km)\\ \hline
$0.00$ & $2.130$ & $11.987$ & $13.497$  \\ \hline
$0.02$ & $2.099$  & $11.774$ & $13.190$ \\ \hline
$0.03$ & $2.031$ & $11.296$ & $12.504$ \\ \hline
$0.04$ & $1.918$ & $10.494$ & $11.380$ \\ \hline
$0.05$ & $1.770$ & $9.655$ & $10.414$ \\ \hline
$0.06$ & $1.604$ & $8.688$ & $9.250$ \\ \hline
\end{tabular}
\end{table}
The thermal evolution mechanism in the further sections is explored for the canonical star with the whole assumed momentum range and the above-listed maximum mass star at the corresponding dark matter momentum. 

This model of dark matter admixed cold neutron star has been adopted previously in the listed literature references to explore its effects on the mass-radius profile, curvature, in-spiral properties and $f$-mode oscillations of the neutron star for various relativistic parameter sets \citep{PhysRevD.99.043016, 10.1093/mnras/staa1435, Das_2021, 10.1093/mnras/stab2387, PhysRevD.104.123006, lourenco2021dark}. However, in this work, we dig the effects of the fermionic dark matter, with neutralino as the suitable dark matter candidate, on the thermal evolution profile of a hot beta equilibrated neutron star. 
\section{Thermal Evolution}\label{ThermEvo}
The energy balance and flux dynamics in terms of relativistic framework to depict the thermal mechanism of a hydrostatic spherically symmetric neutron star were initiated by Thorne. The heat diffusion in the neutron stars within the relativistic framework can be explored using these energy balance and transport equations \citep{1977ApJ...212..825T},   
\begin{eqnarray}
\frac{\partial}{\partial\,r}(Le^{2\phi(r)}) = -\,\frac{4 \pi r^{2} e^{\phi(r)}}{\sqrt{1-\frac{2GM}{c^{2}r}}} \Big(C_{V} \frac{\partial\,T}{\partial\,t} - e^{\phi(r)}(Q_{\nu} + Q_{h}) \Big),
\label{eq5}
\end{eqnarray}
\begin{eqnarray}
\frac{\partial}{\partial\,r}(Te^{\phi(r)}) = -\frac{L}{\kappa\,4\pi r^{2}}\,\frac{e^{\phi(r)}}{\sqrt{1-\frac{2GM}{c^{2}r}}},
\label{eq6}
\end{eqnarray}
where $t$ and $\phi$ are the time coordinate and Schwarzschild gravitational potential respectively. $C_{V}$ is the specific heat, $Q_{\nu}$ represents neutrino emissivity, $Q_{h}$ is associated to the heat production per unit volume due to several factors (for instance, dissipation of rotational energy, magnetic dissipation etc.) \citep{ 1997MNRAS.292..167U, 2000A&A...360.1052P} and ignored in the present calculations, $\kappa$ is the thermal conductivity and $L$ stands for the total luminosity with neutrino and radiation part combined. The solution of these two coupled differential equations endue us with the surface temperature (\,$T_{s} (t)$\,) of the star and the distribution of the internal temperature of the star ($T$) as a function of time. The star's surface temperature in a locally flat reference frame ($T_{s}$) is related to the apparent surface temperature for a distant observer ($T_{s}^{\infty}$) as $T_{s}^{\infty} = T_{s} \sqrt{1 - 2GM/(c^{2}R)}$. The numerical simulation to get the solution of these equations would require some initial and boundary conditions, which are adopted as:  $L_{r}=0$ (no initial heat flux) ; $L(r=R,t)=4\pi \sigma_{S} R^{2} T^{4}_{s} (t)$ (boundary) and $Te^{\phi} (t=0) = T_{0}$ (initial redshiftted temperature), where $\phi$ determines the redshift and depends on the model used. 
\subsection{Neutrino Emissivity}
The exact metric for the total emissivity of a neutron star is a complicated business due to the number of neutrino reactions in the neutron star crust as well as in its core, which can be derived using the electroweak interaction Glashow-Weinberg-Salam theory \citep{BILENKY198273}. These neutrino reactions are the most powerful source and are mainly responsible for the initial cooling stage of the young stars. As we know, the neutron star's crust and core consist of different complex states of matter, so, the neutrino emission in the crust and core is also controlled or dominated by an entirely different mechanism. Since, it is irresolvable to include each neutrino reaction taking place in the stellar objects and also the main aim of the work is to observe the dark matter effects on the resultant emissivity, so, we incorporated a few of them in our present work, most of which are mainly responsible for the major contribution separately in the crust and the core segment of the star. 
\\
(i) Plasmon decay [$\gamma \longrightarrow \nu\bar{\nu}$].
It occur due to the electron-medium interaction and controlled by the electron plasma frequency, which further depends on the number density and the chemical potential of the electron in the defined beta equilibrium matter. This process is mainly effective at lower densities and the adopted mathematical expression for the same can be found in the listed references \citep{PhysRevLett.66.1655, 1992ApJ...395..622I, 10.1111/j.1365-2966.2007.12342.x}.\\
(ii) Pair annihilation [$ e^{-}e^{+} \longrightarrow \nu\bar{\nu}$].
The neutrino emissivity due to electron-proton pair annihilation is not much significant due to the small fraction of positrons in the neutron stars \citep{PhysRevLett.5.573, YAKOVLEV20011, 10.1111/j.1365-2966.2007.12342.x}.
\\
(iii) $e^{-}$-nucleus Bremsstrahlung [\,$e\,\{A,Z\} \longrightarrow e\,\{A,Z\}\,\nu\bar\nu$\,], $\{A,Z\}$ represents the nuclei.
The neutrino emissivity due to $e^{-}$-nucleus Bremsstrahlung processes is affected by the shape of the nuclei in the crust of the star and is considered as spherical for the present calculations. For detailed analytical discussion and adopted mathematical expression, the readers are advised to go through these references  \citep{1973ApJ...180..911F, 1996A&A...314..328H, 1999A&A...343.1009K, YAKOVLEV20011}.\\
(iv) Neutron-nucleus Bremsstrahlung [\,$n\,\{A,Z\} \longrightarrow n\,\{A,Z\}\,\nu\bar\nu$\,].
This process is operative in the crustal envelope of the neutron star and enhance the neutrino emissivity magnitude of crust in a considerate manner. The neutrino emissivity expression for the non-superfluid stellar matter due to Neutron-nucleus Bremsstrahlung processes is borrowed from \citep{1977Ap&SS..48..159F}.
\\
(v) Neutron-neutron Bremsstrahlung [$nn \longrightarrow nn\,\nu\bar\nu$];\, proton-proton Bremsstrahlung [$pp \longrightarrow pp\,\nu\bar\nu$];\, neutron-proton Bremsstrahlung [$np \longrightarrow np\,\nu\bar\nu$];
\begin{eqnarray}
Q^{ij} = 1.5\times10^{20}\, \left(\frac{M_{i}^{\ast}}{M_{i}}\right)^{2}\left(\frac{M_{j}^{\ast}}{M_{j}}\right)^{2}\left(\frac{n_{j}}{n_{0}}\right)^{\frac{1}{3}}\,\alpha_{ij}\,\beta_{ij}\mathcal{N}_{\nu}\left(\frac{T}{10^{9}}\right)^{8},
\label{eq11}
\end{eqnarray}
where, $\mathcal{N_{\nu}} \xrightarrow{}$ neutrino flavour number ($3$). The pair $(ij)$ is defined as the combination of neutron-neutron (nn), proton-proton (pp) and neutron-proton (np), for which the corresponding values of the parameters $\alpha_{ij}$ and $\beta_{ij}$ are: $\alpha_{nn} = 2.95$,  $\beta_{nn} = 0.56$;\, $\alpha_{pp} = 0.55$,  $\beta_{pp} = 0.7$;\, $\alpha_{np} = 1.06$,  $\beta_{np} = 0.66$ \citep{1979ApJ...232..541F}. These processes are the moderate sources for the neutrino emissions in the stellar cores and also affected considerably in presence of neutron superfluidity .\\
(vi) Coulomb Bremsstrahlung [\,$ep \longrightarrow ep\,\nu\bar\nu$, $\mu p \longrightarrow \mu p\,\nu\bar\nu$, $ee \longrightarrow ee\,\nu\bar\nu$, $\mu\mu \longrightarrow \mu\mu\,\nu\bar\nu$, $e\mu \longrightarrow e\mu\,\nu\bar\nu$\,].
These processes are similar to nucleon Bremsstrahlung but they originate due to Coulomb forces instead of strong collisions, as in case of baryon Bremsstrahlung and the mathematical expression for neutrino emissivities due to the Coulomb Bremsstrahlung are borrowed from \citep{ 1999A&A...343.1009K}.
\begin{figure}
\centering
\includegraphics[width=1\columnwidth]{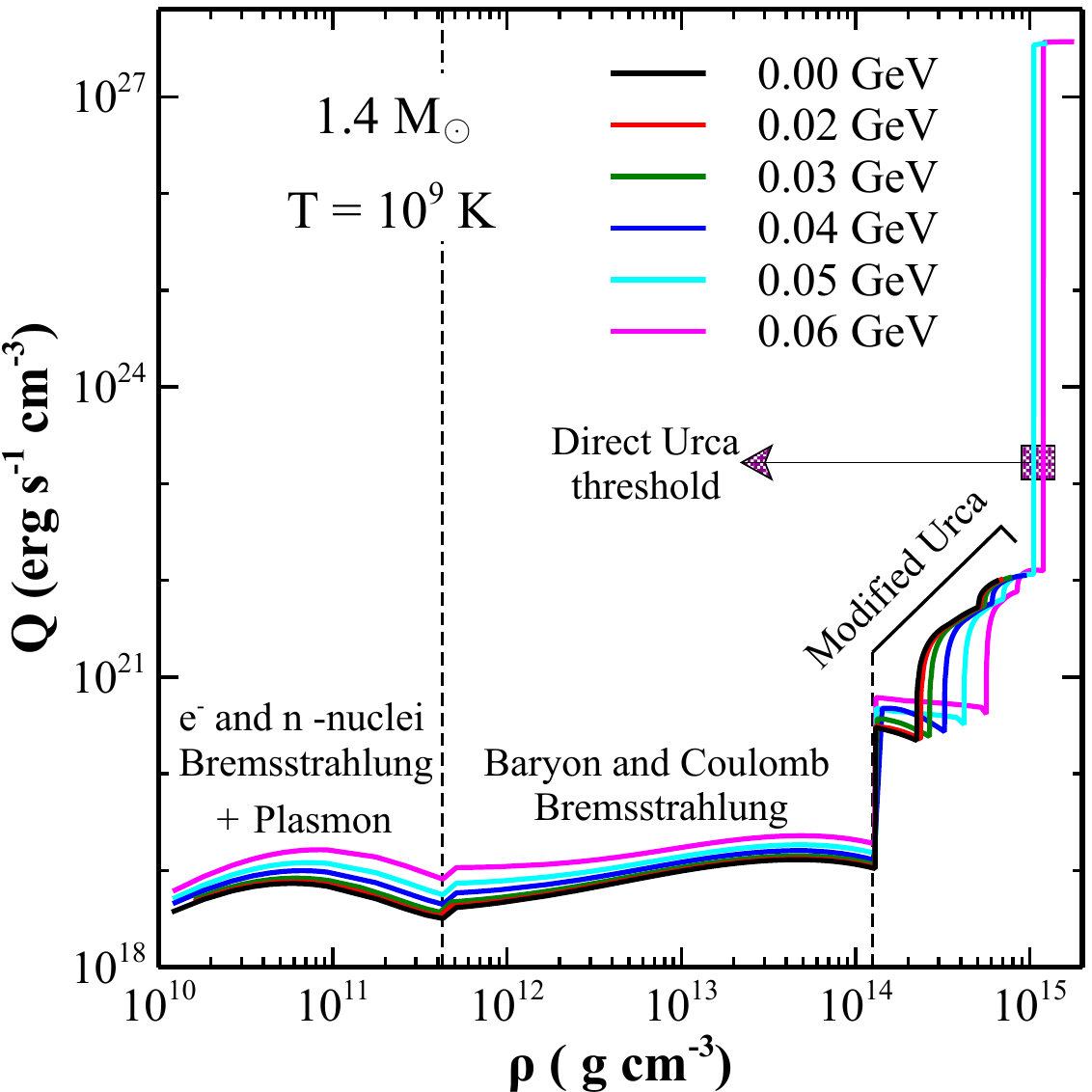}
\caption{Neutrino emissivity of a canonical star as a function of mass density at $T = 10^{9}\, K$ for different configuration of dark matter momentum. The black colour curve ($k_f^{DM}=0.00$ GeV) represents the profile of a canonical star obtained using IOPB-I parameter set without any dark matter admixture.}
\label{fig2}
\end{figure}
\\
(vii) Direct Urca processes [\,$n \longrightarrow pe^{-}\bar\nu$, \,$pe^{-} \longrightarrow n\nu$, \,$n \longrightarrow p\mu\bar\nu$, \,$p\mu \longrightarrow n\nu$\,];
\begin{eqnarray}
Q^{DU} = \frac{457\pi}{10080}\,G_{F}^{2}\,C^{2}\left(1+3g_{A}^{2}\right)\frac{M_{n}^{\ast}\,M_{p}^{\ast}\,m_{e}^{\ast}}{\hbar^{10}\,c^{3}}\,(k_{B}\,T)^{6}\,\Theta_{npe},
\label{eq14}
\end{eqnarray}
where $G_{F}$ and $g_{A}$ are the Fermi weak and axial coupling constants respectively, $C = 0.973$ is known as Cabibbo factor and $\Theta_{npe}$ is the step function which describes the onset of direct Urca process inside the core of neutron star. The value of $\Theta_{npe}$ is unity only if the triangle inequality between the Fermi momentum of neutrons, protons and electrons is satisfied i.e. $(k_{F_{n}}-k_{F_{p}}-k_{F_{e}})\leq0$, otherwise it is zero. For direct Urca reactions involving muons, the expression for the neutrino emissivity remains the same except the step function, which in case of muons will be read as $\Theta_{np\mu}$. The threshold density for the muon Urca processes is higher than that of electron involving events. Once the threshold boundary density for the direct Urca to operate is reached, it dominates all other neutrino processes significantly and responsible for the very fast cooling of the dense cores. \citep{PhysRev.59.539, PhysRevLett.66.2701}. 
\\
(viii) Modified Urca processes \\ 
(a)\, (\,Neutron branch\,)\, [\,$nn \longrightarrow npe^{-}\bar\nu$, $nn \longrightarrow np\mu\bar\nu$, $npe^{-} \longrightarrow nn\nu$, $np\mu \longrightarrow nn\nu$\,] 
\begin{eqnarray}
Q^{Mne} = 6.224\times10^{21}\left(\frac{M_{n}^{\ast}}{M_{n}}\right)^{3}\left(\frac{M_{p}^{\ast}}{M_{p}}\right)\left(\frac{n_{p}}{n_{0}}\right)^{1/3}\left(\frac{T}{10^{9}}\right)^{8}.
\label{eq15}
\end{eqnarray}
\begin{figure}
\centering
\includegraphics[width=1\columnwidth]{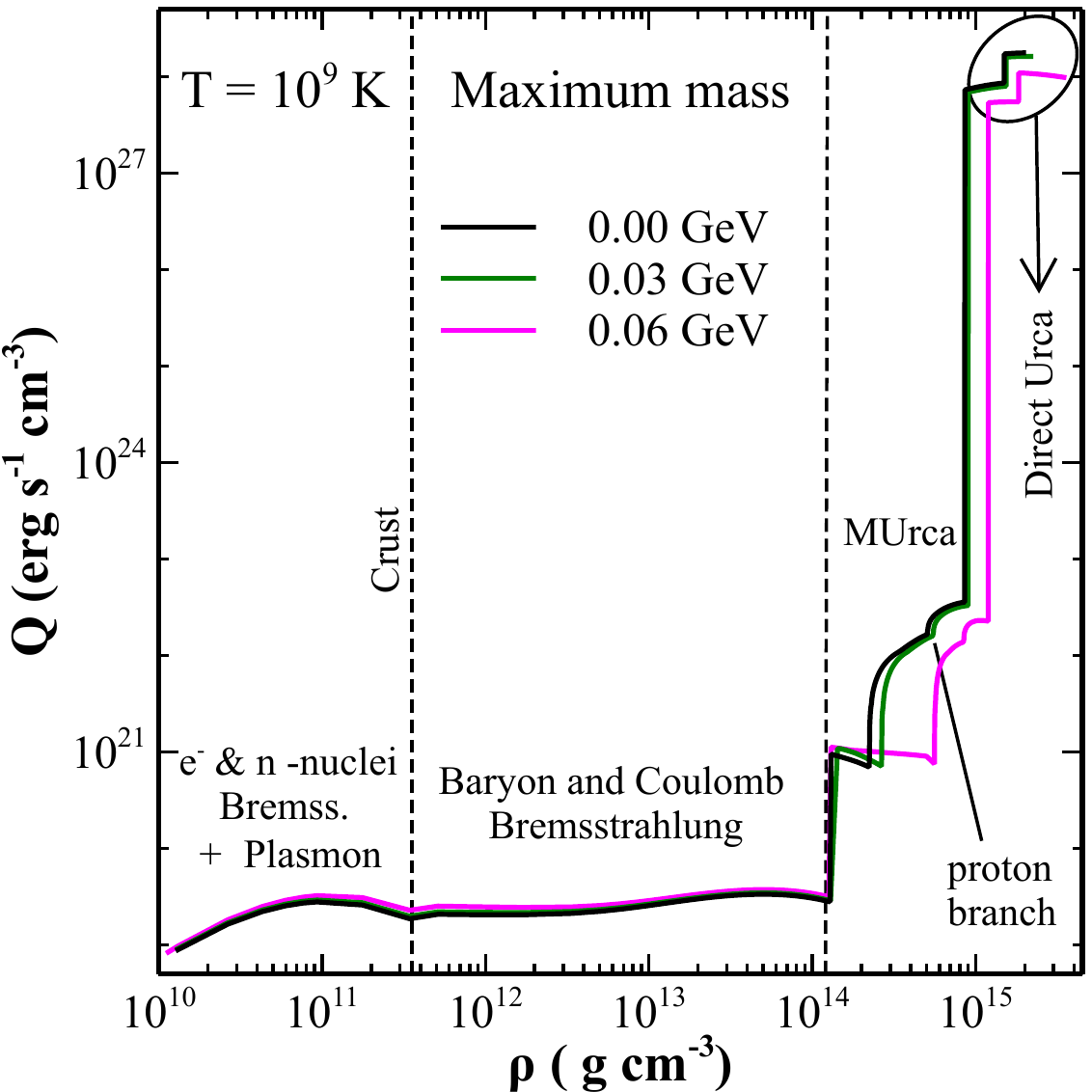}
\caption{Neutrino emissivity is shown for the maximum mass profile of the star acquired with the corresponding dark matter momentum. The obtained maximum mass of the star for $k_{f}^{\rm DM} = 0.03$ and $0.06$ GeV are $2.031$ and $1.604 \, M_{\odot}$ respectively.  }
\label{fig3}
\end{figure}
(b)\, (\,Proton branch\,)\, [\,$pn \longrightarrow ppe^{-}\bar\nu$, $pn \longrightarrow pp\mu\bar\nu$, $ppe^{-} \longrightarrow pn\nu$, $pp\mu \longrightarrow pn\nu$\,]
\begin{eqnarray}
Q^{Mpe} = \left(\frac{M_{p}^{\ast}}{M_{n}^{\ast}}\right)^{2} \frac{\left(k_{F_{e}}+3k_{F_{p}}-k_{F_{n}}\right)^{2}}{8k_{F_{e}}k_{F_{p}}}Q^{Mne}\Theta_{Mpe}.
\label{eq16}
\end{eqnarray}
The proton branch of the modified Urca process require a threshold density to operate, which is controlled by the theta function ($\Theta_{Mpe}$) and is non-zero for $k_{F_{n}}\,<\,3k_{F{p}}+k_{F_{e}}$. In the case of muon involving cycles which depends on the appearance of muons inside the stars, both the emissivity expressions should be multiplied by $(n_{\mu}/n_{e})^{1/3}$ and the threshold for the proton branch is determined by $k_{F_{n}}\,<\,3k_{F{p}}+k_{F_{\mu}}$. The reaction mechanism and neutrino production rate of this process is different from that of the direct Urca processes. It requires an additional nucleon for the conservation of momentum, has a slower reaction rate and, happens to be inoperative in the presence of direct Urca reactions \citep{PhysRevLett.12.413, 1979ApJ...232..541F, 1995A&A...297..717Y, YAKOVLEV20011}.

All the above-introduced processes circumstantially depend on the superfluid effects, high magnetic field, presence of hyperons, condensation of pions or kaons and a high dense quark core of the neutron stars. The inspection of neutrino emission rate from the neutron stars along with the inclusion of all these factors is beyond the scope of this work and may be adopted later in different context for the forthcoming projects. The keen readers are advised to go through the corresponding listed references, mainly \citep{YAKOVLEV20011} \& \citep{Potekhin2015}, for more detailed derivations and explicit discussions of each neutrino emission process along with the superfluid effects. The effects of the dark matter component on the neutrino emission processes in a canonical and maximum mass star are depicted in figs. \ref{fig2} and \ref{fig3}. The thermal contour of the neutron star in the early stage of the star is controlled by the outermost crust of the star, which in turn is mostly influenced by the neutrino emissivity due to plasmon decay  \citep{1987ApJ...312..711N, 10.1046/j.1365-8711.2001.04323.x}. The increment of the dark matter factor in a coequal mass neutron star amplifies the plasma frequency by means of electron Fermi momentum and develops the neutrino sources in the outer layers of the star, which results in the increased neutrino emissivity (fig. \ref{fig2}). Since, the initial age cooling of a neutron star is independent of what is happening in the core and the crustal part is the prime source of neutrino emissions, it turns out that a dark matter admixed neutron star cools more expeditiously in formative years. The compositional profile of a pure canonical neutron star (i.e. without any dark matter admixture) obtained using the IOPB-I parameter set does not support the direct Urca process and the neutrino emissivity through modified Urca processes is responsible for the fastest cooling scenario of such stars. The activation of the proton branch for electron and muon case takes place separately depending upon the fulfillment of triangle inequality by Fermi momenta of the particles and the activation point as pointed in figs. \ref{fig2} and \ref{fig3} in the form of bumps within the modified Urca dominance region. The proton branch enhances the emissivity of the modified Urca process by a factor of ~10 and the activation density is higher for a neutron star with greater dark matter composition. However, we observe that if the theoretically considered availability of dark matter segment in such stars is such that $k_{f}^{\rm DM}>0.04$ GeV, then the compositional abundance of the star enables the direct Urca reactions in the core (fig. \ref{fig2}). The dark matter admixture increases the central density of the star, as shown in fig. \ref{fig1}, which is accompanied by an increase of proton and lepton fraction. With the high fraction of protons and leptons in such dark matter admixed canonical stars ($k_{f}^{\rm DM}>0.04$ GeV), the inevitable triangle inequality for direct Urca process is satisfied and an additional highly efficient neutrino emission channel is operative in such stars. The metric for the neutrino emissivity of the direct Urca process is almost $10^{5}$ times higher than the other neutrino emission processes, so navigating the threshold density for direct Urca reaction results in a very rapid cooling of the canonical star. The turning on density for direct Urca processes of DM admixed canonical neutron star in case of $k_{f}^{\rm DM}=0.05$ and $0.06$ GeV are $1.052\times10^{15}$ and $1.195\times10^{15}$ g\,cm$^{-3}$ respectively. The direct Urca neutrino emissions are present in both cases of the maximum mass star i.e. with or without dark matter component and the triangle inequality for muons is also satisfied in the case of maximum mass star (fig. \ref{fig3}), which opens an extra channel for neutrino production and further enhance the neutrino emissivity. The electron and muon channel threshold density of direct Urca process in the maximum mass scenario of a pure neutron star ($k_{f}^{\rm DM}=0.00$ GeV) with IOPB-I parameter set are $8.585$ and $14.974$ ($10^{14}$ g \,cm$^{-3}$) respectively. The threshold value of density for DM admixed maximum mass neutron star with $k_{f}^{\rm DM}=0.06$ GeV is same as that of a canonical star. 
\subsection{Heat Capacity and Thermal Conductivity}
The energy required to increase the temperature of the star is theoretically studied in terms of the specific heat capacity, measured by summing the individual contribution of heat from all the considered particles in the neutron star model. The total heat capacity of a neutron star in the crustal part is dominated by the relativistic electron degenerate gas and the lattice of atomic nuclei present in different phases of the neutron star's crust. The contribution in the core part of the star received from the neutrons, protons, electrons and muons for the present case and other exotic particles like hyperons, quarks etc., which are not considered in the present model. The theoretical mechanism for the specific heat capacity generated in the crust part of the star due to the coulomb ion vibrations in the lattice depends on anharmonicity, phase transition, Debye frequency, order parameter ($\Gamma$) etc. \citep{PhysRev.122.1437, PhysRevA.26.2255}. The final adopted mathematical expressions can be found in the reference \citep{1991ApJS...75..449V}, and the interested readers for detailed procedure of derivations are advised to go through the listed references \citep{PhysRevE.64.057402, 2010CoPP...50...82P}.

An updated and generally used expressions for the specific heat of the ions in star's crust are derived by Haensel et. al. and advised to the readers for relevant simulations in future projects \citep{2018A&A...609A..74P}. Apart from the ions, the second component that majorly contributes to the specific heat of the crust is the relativistic degenerate electron gas.
\begin{figure}
\centering
\includegraphics[width=1\columnwidth]{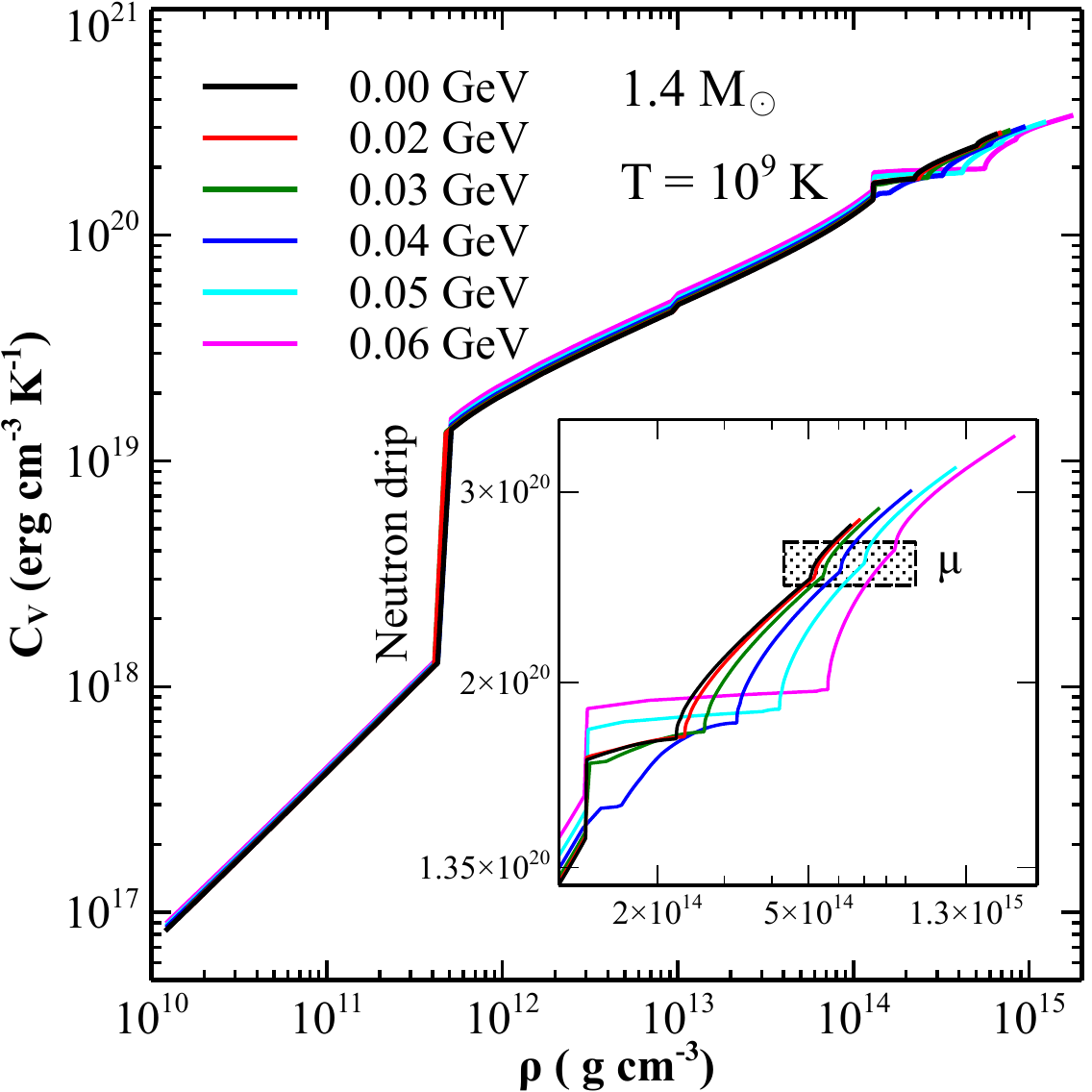}
\caption{Specific heat of a dark matter admixed canonical star as the function of density at $10^{9}\,K$ temperature. Dark matter momentum is varied from $0.00$ GeV to $0.06$ GeV. The stellar core portion is depicted in the inset. The region of the appearance of muons is incited in rectangular box.}
\label{fig4}
\end{figure}
For temperature of the interest in the present case $T << \mu_{i}$ and in the absence of superfluidity, the specific heat of each constituent particle can be stated as,
\begin{eqnarray}
C_{V} &=& \sum_{i} \frac{M_{i}^{\ast}\,n_{i}}{k_{F_{i}}^{2}}\,\pi^{2}\,k_{B}^{2}\,T,
\label{eq21}
\end{eqnarray}
where $i$ sums over neutrons, protons, electrons and muons \citep{Page_2004, 2015SSRv..191..239P}. The effective mass of leptons (electron and muon) is defined as $\mu_{e,\mu}/c^{2}$, while for nucleons (proton and neutron) it is the in-medium effective mass acquired for the IOPB-I parameter set in the presence of dark matter as defined in section \ref{EoS}. The cumulative specific heat of a dark matter admixed canonical star and maximum mass star profiles for IOPB-I parameter set is depicted in figs. \ref{fig4} and \ref{fig5} respectively at $T = 10^{9} K$. The influence of dark matter on the heat capacity of the star is embodied only through the interference of dark matter particles on the effective mass and compositional distribution of the constituent particles. As for the crustal segment, the major contribution to the specific heat is through the ionic oscillations and degenerate electron gas which is not much influenced by the dark matter contribution and depends on the shape, which is considered spherical in the present case, structure and nature of the envelope considered (heavy or light element), so, there is no influential change is observed in the specific heat with dark matter variation at low density. With the increase in density, neutron drips out of the nuclei and sharp enhancement in the heat capacity is recorded at the boundary. In the high dense region (core of the star), the effective mass and abundance of the nucleons and leptons, mainly neutrons, is influenced considerably by the involvement of dark matter contribution which is reflected on the specific heat of the star too. The increase in the dark matter momentum hints for the suppression of nucleon's effective mass and number density product in the core which results in the reduction of heat capacity for a larger value of $k_{f}^{\rm DM}$ at a certain density. The appearance of the muons which is controlled by the beta equilibrium conditions of the star results in the further increment of the heat capacity inside the core. It should be noted that, however this algorithm scheme predicts the consistent pattern about specific heat capacity of the stellar core but is largely uncertain due to the absence of cooper pairing of neutrons, magnetic field, quark degrees of freedom and superfluidity.

Another important ingredient required to unfold the thermal dynamics of young non-relaxed neutron stars is thermal conductivity. Thermal conduction is also extremely responsible for the internal thermal relaxation of a neutron star.
\begin{figure}
\centering
\includegraphics[width=1\columnwidth]{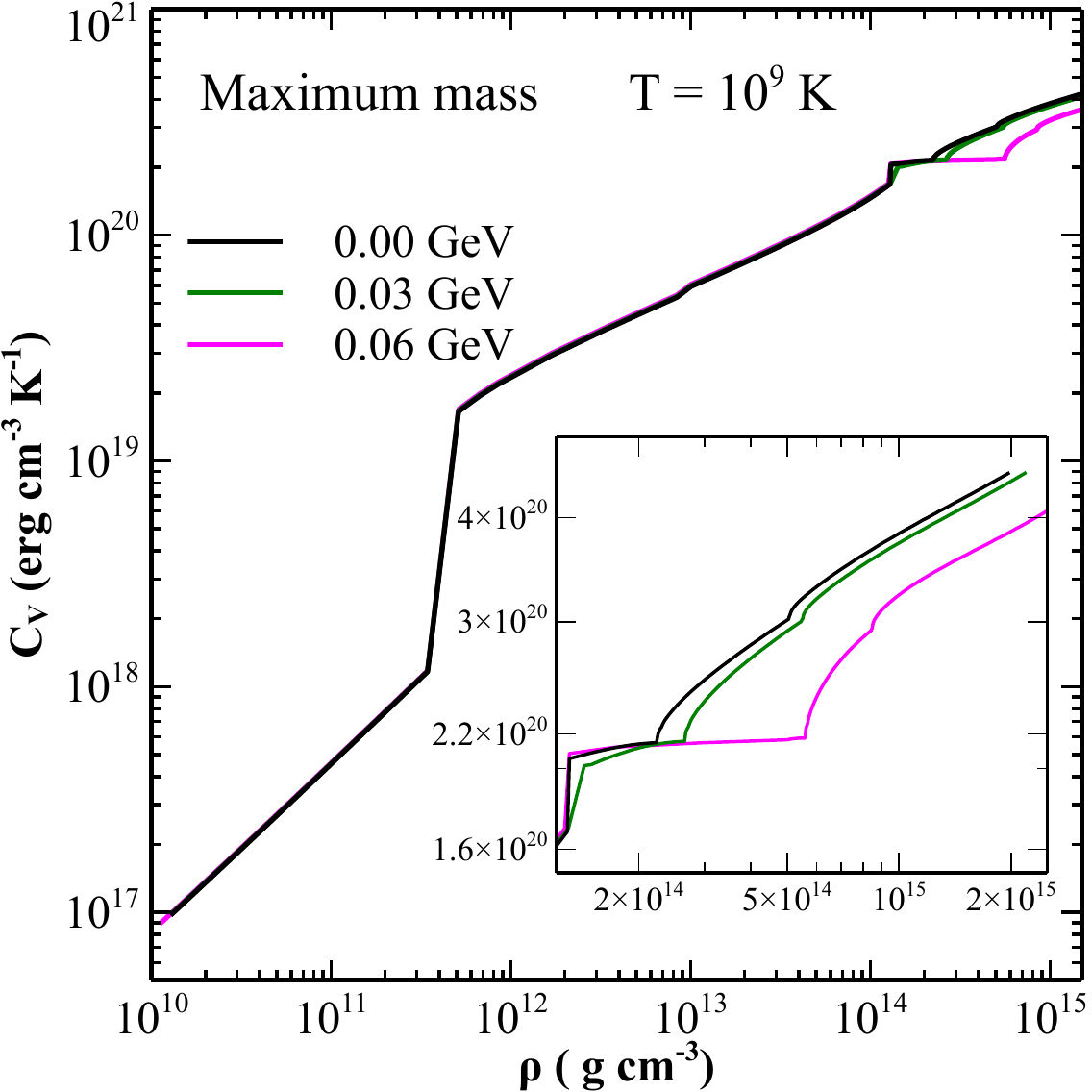}
\caption{Same as Fig. \ref{fig4}, but for the profile of maximum mass star obtained for different values of $k_{f}^{\rm DM}$ with IOPB-I parameter set as apprised in section \ref{EoS}.}
\label{fig5}
\end{figure}
The radiative luminosity of the stars is determined by the mechanism of heat transfer in the outermost layers and the ocean, which is carried out mostly by electrons. In the absence of magnetic field the heat transfer through the phonons (ionic thermal conductivity) is not of much importance. So, we mainly consider the electron heat conduction in outer crust  of the neutron star. The expression for the electron thermal conductivity coefficient in the outer envelopes with multiphonon phenomenon and long-range nucleus correlations in Coulomb solid and liquid of nuclei had been derived explicitly by Potekhin and Yakovlev, and is given by \citep{2001A&A...374..213P, Potekhin2015}
\begin{eqnarray}
\kappa_{e} &=& \frac{\pi^{2}\,k_{B}^{2}\,n_{e}\,T}{3\,m_{e}^{\ast}\,\nu_{e}},
\label{eq22}
\end{eqnarray}
where, $m_{e}^{\ast}$ is the relativistic kinematic electron mass and $\nu_{e}$ is the effective collision rate due to electron-ion ($\nu_{ei}$) and electron-electron ($\nu_{ee}$) scattering. The kinematic electron mass is defined as $m_{e}^{\ast} = \sqrt{1 + x_{r}^{2}}$ with $x_{r} = \Big(\frac{1.027\,\rho\,Z}{10^{6}\,A}\Big)^{1/3}$. The effective collision frequency in the outermost crust due to the electron-electron and electron-ion components can be simply add up in accordance with the Matthiessen rule and depends on the phase state and composition of the nuclei and ions adopted in the crustal part. 
\begin{figure}
\centering
\includegraphics[width=1\columnwidth]{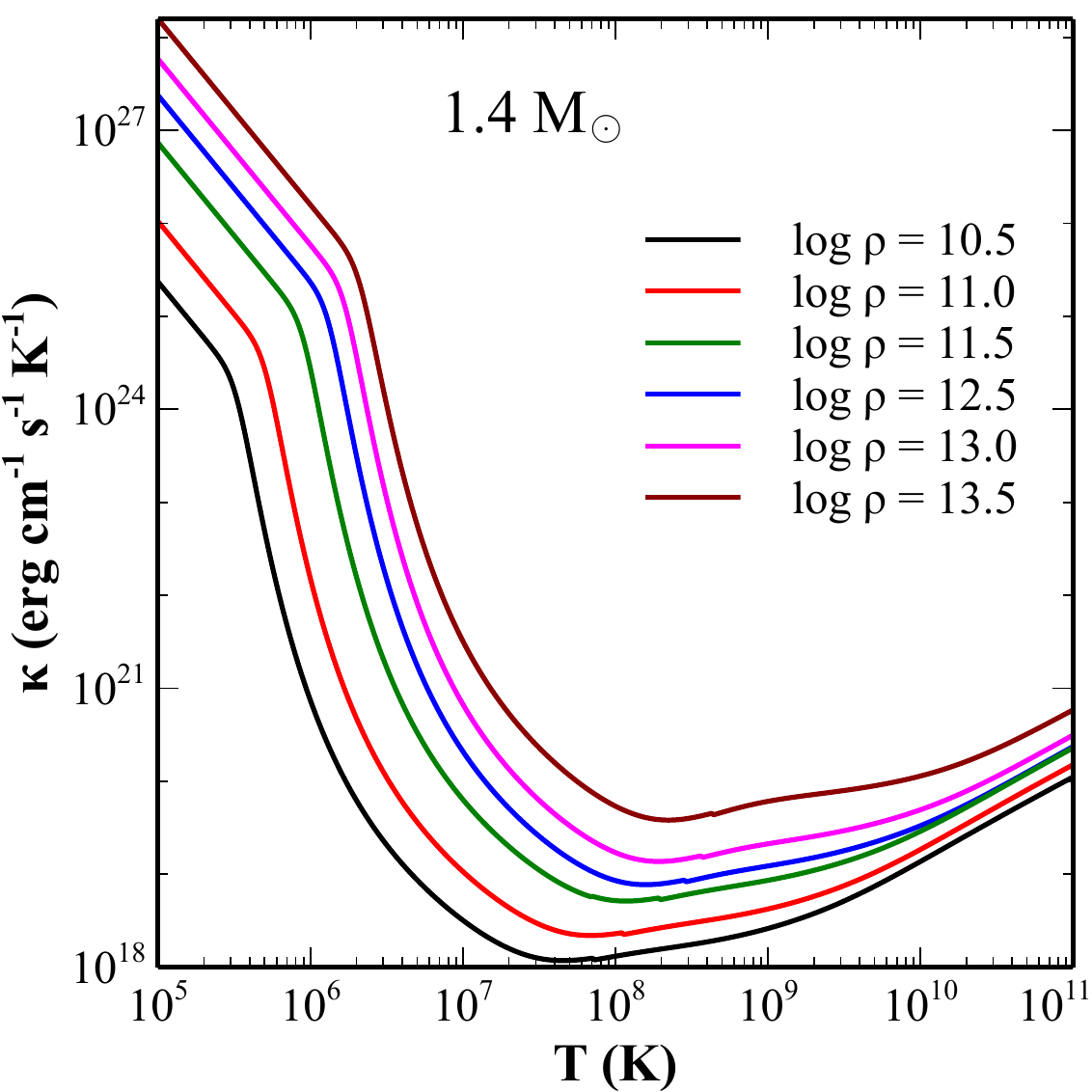}
\caption{Thermal conductivity for a pure canonical star ($k_{f}^{\rm DM} = 0.00$ GeV) with IOPB-I parameter set as a function of temperature at different values of density. Mainly outer and inner crust is shown here.}
\label{fig6}
\end{figure}
The electron-ion collision frequency in case of degenerate and non-degenerate electron gas for Coulomb solid and ion liquid regime was first proposed by Yakovlev and Urpin \citep{1980SvA....24..303Y}, which later had been unified in a single regime by Potekhin {\it et al.} and adopted in the present work for simulations. Accordingly, the electron-electron scattering in the stellar crust like environment, first entertained by Flowers and Itoh at small temperatures \citep{1976ApJ...206..218F}, and then undergo several modifications with the inclusion of charge-current interactions, Landau damping, magnetic field etc. by various authors and the final admittable expression of $\nu_{ee}$ for the present simulations is stated in the same work by Potekhin {\it et al.} as mentioned above \citep{Potekhin1999TransportPO}. For the intermediate density (inner crust), the thermal conductivity due to neutrons also comes into play. Also, the electron conductivity will be affected by the shape and size of the nuclei in the crust and one should include the form factor modifications in the collision frequency calculations \citep{10.1046/j.1365-8711.2001.04359.x}. The thermal conduction coefficient of neutrons in that region can be estimated using the same equation as that for electron (eq. \ref{eq22}) by replacing the electron number density ($n_{e}$) and kinematic mass ($m_{e}^{\ast}$) with the corresponding neutron number density ($n_{n}$) and the neutron in-medium effective mass ($M_{n}^{\ast}$) as defined in section \ref{EoS}. The effective neutron collision frequency ($\nu_{n}$) for the said region can also be acquired in the same way by adding the neutron-ion ($\nu_{ni}$) and neutron-neutron collision ($\nu_{nn}$) counterparts. For the case of degenerate neutrons and in the negligence of ion correlated collisions, the neutron-ion collision rate will be represented by,
\begin{eqnarray}
\nu_{ni} &=& \frac{\pi\,n_{i}\,R_{n}^{2}\,k_{F_{n}}\,M_{n}^{\ast}}{\sqrt{M_{n}^{\ast\,2}c^{2} + k_{F_{n}}^{2}}},
\label{eq23}
\end{eqnarray}
where $n_{i}$ is the ion number density and $R_{n}$ is the neutron radius of atomic nuclei in the crust  \citep{Bisnovatyi-Kogan1982, 2013A&A...560A..48P}. The neutron-neutron collision frequency as a function of temperature and effective self scattering cross-section ($S_{nn}$) in the stellar medium evaluated by Shternin {\it et al.} as \citep{PhysRevC.88.065803},
\begin{eqnarray}
\nu_{nn} &=& \frac{12.8\,k_{B}^{2}\,T^{2}\,}{\hbar^{3}}\,\left(\frac{M_{n}^{\ast\,3}}{M_{n}^{2}}\right)\,S_{nn}.
\label{eq24}
\end{eqnarray}
The mechanism for the in-medium cross-section was later revisited by Baiko {\it et al.} using the Bonn potential and provided in the form of simple analytical expression \citep{2001A&A...374..151B}.
\begin{figure}
\centering
\includegraphics[width=1\columnwidth]{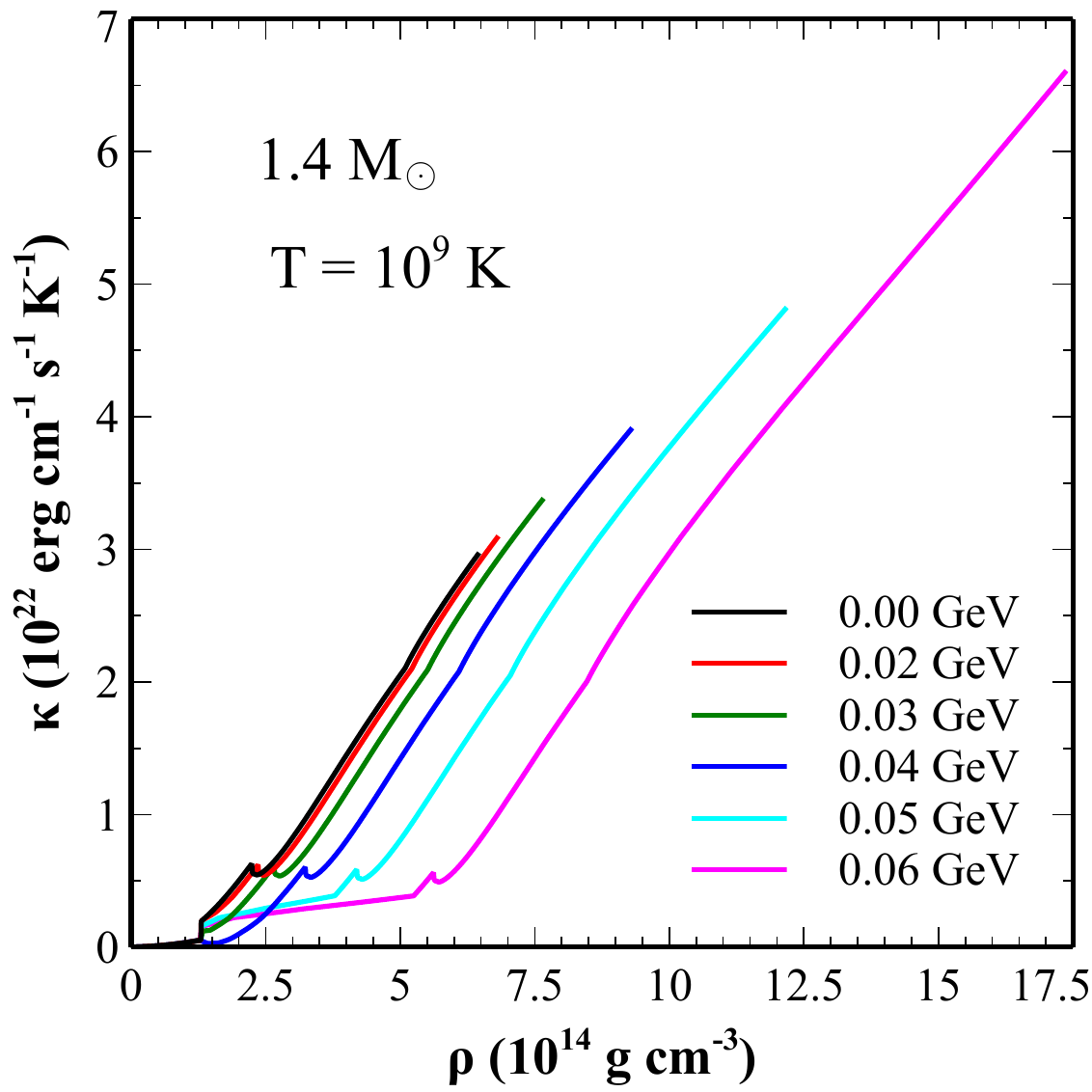}
\caption{Total thermal conductivity profile of a dark matter admixed canonical star with IOPB-I parameter set as a function of neutron star's density. Here thermal conductivity is shown for a fixed temperature ($10^{9}\, K$).}
\label{fig7}
\end{figure}
Total thermal conductivity of a canonical star due to electron and neutron contribution for the considered crustal EOS \citep{Chamel2008} is depicted in fig. \ref{fig6} within the temperature range $10^{5} - 10^{9}\,K$. High thermal conductivity results in more rapid cooling scenarios, so, once the temperature around $10^{8}\,K$ is achieved by the stellar crust, the smaller thermal conductivity indicates the slow cooling era of the neutron star. The thermal conductivity of electrons in the outer envelopes of the star is suppressed up to several orders of magnitude due to the finite size atomic nuclei scatterers and the scattering due to charged impurities is neglected in the present calculations. Also, at a constant temperature the thermal conductivity of the crustal part increases with an increase in density and we observe that this effect is reversed in the core with an increment of the dark matter percentage, as shown in fig. \ref{fig7}.

The heat transport in the stellar core is mainly categorized into two parts: baryon heat conduction and lepton heat conduction. The total thermal conductivity magnitude is the result of the solution of the multi-component system of transport equations and the analytical expression for each type of particle ($i$) is analogous to equation (\ref{eq22}), 
\begin{eqnarray}
\kappa_{i} &=& \frac{\pi^{2}\,k_{B}^{2}\,n_{i}\,T\,\tau_{i}}{3\,m_{i}^{\ast}},
\label{eq25}
\end{eqnarray}
where $i = n, p, e^{-} \,\&\, \mu$ and the other variables have their usual meaning as defined above. It should be noted here that for core simulations the effective mass of leptons is determined by their dynamical chemical potential which is controlled by the beta equilibrium condition for the stellar core. The effective mass of the particle and the corresponding number density is acquired using mean-field formalism for IOPB-I parameter set with different dark matter momenta, as discussed in section \ref{EoS}. The formal solution of effective relaxation times for baryons within the high dense environment using the differential collision probability and solving the set of algebraic equations are obtained by Shternin {\it et al.} as \citep{PhysRevD.78.063006, PhysRevC.88.065803},
\begin{eqnarray}
\sum_{i,j=\{p,n\}} \frac{64\, M_{i}^{\ast} M_{j}^{\ast\,2}\,T^{2}}{5\, M_{N}^{2}\, \hbar^{3}}\,S_{ij}\,\tau_{j} = 1. 
\label{eq26}
\end{eqnarray}
The detailed calculations for the effective cross-section ($S_{ij}$) in beta equilibrated matter are performed in \citep{2001A&A...374..151B} and borrowed from the same for present calculations. Coulomb potential is the main mechanism for scattering in the case of leptons, and the effective relaxation time for electrons and muons will be given by \citep{1995NuPhA.582..697G},
\begin{eqnarray}
\tau_{e} = \frac{\nu_{\mu} - \nu_{e \mu}^{'}}{\nu_{e} \nu_{\mu} -  \nu_{e \mu}^{'}  \nu_{\mu e}^{'}};\,\,\,\,\,\,\,\,\,\,\,\,\,\,\,\,
\tau_{\mu} = \frac{\nu_{e} - \nu_{\mu e}^{'}}{\nu_{e} \nu_{\mu} -  \nu_{e \mu}^{'}  \nu_{\mu e}^{'}},
\label{eq27}
\end{eqnarray}
with $\nu_{i} = \sum_{j} \nu_{ij}$; where $i$ denotes the leptons ($e\,\&\,\mu$) and $j$ stands for all the considered charged particles in the star core ($p, e, \mu$). The $\nu_{\mu e}^{'}$ and $\nu_{e \mu}^{'}$ are crossed collision frequencies that coupled the heat transport of electrons and muons. The total self and crossed collision frequencies for electrons and muons in the ultra-dense cores are calculated by solving the multidimensional integral over colliding particles with Coulomb screening momentum ($q_{0}$) taken into account and are expressed in the form of simple analytical expressions \citep{GNEDIN1995697},
\begin{eqnarray}
\nu_{ii} &=& \frac{5}{2} \frac{\beta_{i}}{\mu_{i}^{2}}\,\,\,\,\,\,\,\,\,\,\,\,\,\, (i = e, \mu), \\
\nu_{ip} &=& \beta_{i}\, \frac{M_{p}^{\ast\,2}}{k_{F_{i}}^{4}}\,\,\,\,\,\,\, (i = e, \mu), \\
\nu_{ij} &=& \frac{\mu_{ip}}{M_{p}^{\ast\,2}} \frac{\nu_{j}^{2}}{c^{4}} \Bigg[1 + \Big( \frac{k_{F_{i}}k_{F_{j}}c^{2}}{4\mu_{i}\mu_{j}}\Big)^{2}\Bigg] (i, j = e, \mu\,;\, i\neq j), \\
\nu^{'}_{ij} &=& \frac{\beta_{i}}{k_{F_{i}} k_{F_{j}} c^{2}}\,\,\,\,\,\,\,\,\,\,\,\, (i, j = e, \mu\,;\, i\neq j),\\
\text{with}\nonumber \\
\beta_{i} &=& \frac{4\pi^{2}}{5\hbar} \Bigg(\frac{e^{2}}{\hbar c}\Bigg)^{2} \Bigg(\frac{k_{F_{i}}}{q_{0}}\Bigg)^{3} \mu_{i} (k_{B}T)^{2} \,\,\,\,\,\,\, (i = e, \mu). \nonumber
\label{eq28}
\end{eqnarray}
We presented here all the possible components of collision frequency with proton being the only charged nucleon, however, in presence of other exotic charged particles (baryon octet or deconfined quarks) there may also be other components, whose impressions on the thermal transport can be explored in future. It should also be noted here that we neglected the possible effects of superfluidity and magnetic field for the time being. After implementing all the above-listed expressions with proper factors and extracted variables from the RMF formalism together, the results for the thermal conductivity of a canonical star with varied dark matter momentum are illustrated in fig. \ref{fig7}. The thermal conductivity curve reflects the same nature as that of EOS for the canonical star with varying dark matter proportions. The appearance of muons in the stellar core further enhances the metric for conductivity but does not seem to be much influential as in the case of neutrino emissivity and specific heat. The magnitude of thermal conductivity, in general, increases as we move from the surface to the core of the star. However, for the high dense portion, the increment in the dark matter percentage at a fixed density reduces the thermal conduction of the star. On the other hand, we also observe that the core of a dark matter concentrated canonical star is highly dense and much more conductive than the no dark matter admixed star. In our calculations, we noticed that at a certain ultra dense core site of dark matter admixed star the thermal transport is greater for higher temperature, which from the stated mathematical expressions can be alluded as the suppression of high-temperature collision effects by the higher momentum metric of nucleons for the core. 

All the above three examined properties (neutrino emissivity, specific heat and thermal conductivity) are required for the numerical simulations related to cooling dynamics and thermal energy transport of young neutron stars, which are performed in the next section.
\subsection{Cooling and Internal Thermal Equilibrium}
\begin{figure}
\includegraphics[width=1\columnwidth]{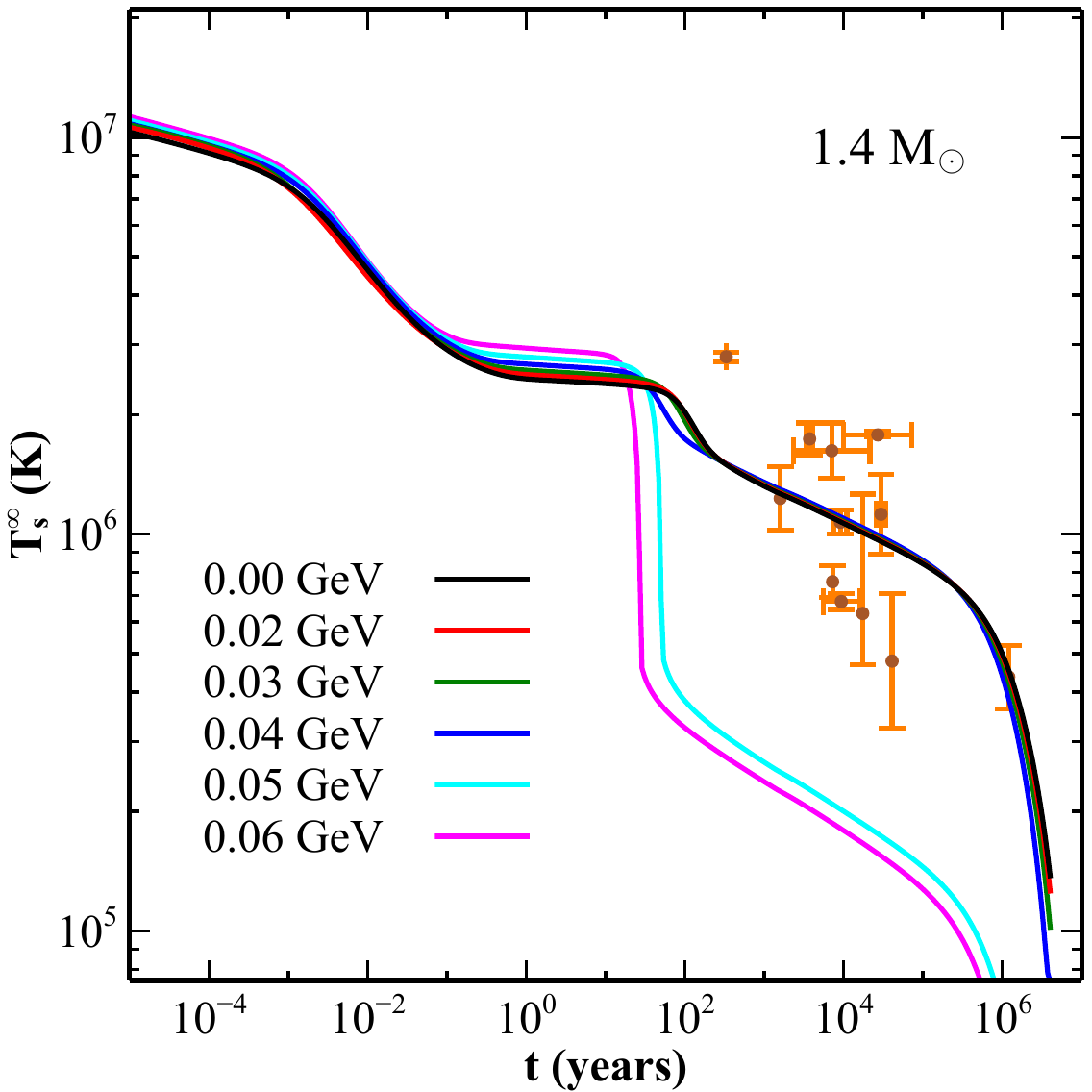}
\caption{The theoretically calculated curves of surface temperatures for a dark matter admixed canonical star as function of stellar age. The observational data of several thermal emitting sources i.e. Cas A NS \citep{Ho_2009}, PSR J1119-6127 \citep{Safi_Harb_2008}, RX J0822-4247 \citep{Zavlin_1999}, 1E 1207.4-5209 \citep{Pavlov_2002}, PSR J1357–6429 \citep{Zavlin_2007}, RX J0002+6246 \citep{Page_2004}, PSR B0833-45 \citep{Pavlov_2001}, PSR B1706-44 \citep{McGowan_2004}, PSR J0538+2817 \citep{Zavlin_2004}, PSR B2334+61 \citep{McGowan_2006}, PSR J0243+2740 \citep{Zavlinbook_2007}, XMMU J1732-344 \citep{refId0} had also been illustrated with upper and lower bounds of surface temperature and age.}
\label{fig8}
\end{figure}
The computed values of $Q$, $C_{v}$ and $\kappa$ are employed in eqs. \ref{eq5} and \ref{eq6} to obtain the surface and internal temperature profiles of the canonical and maximum mass stars. Some of the earlier works which provide the solution algorithm to these equations are \citep{1979ApJS...39...29R, 1982ApJ...255..624R, 1984ApJ...278..813F, 1991ApJS...75..449V}, later developed through finite-difference Euler and Henyey methods by several researchers in various works related to cooling dynamics of neutron star   \citep{10.1046/j.1365-8711.2001.04359.x, 2019A&A...629A..88P}. The key point to solve these equations is to divide the spatial points in infinitesimal spherical layers of thickness $dr_{i}$ ($\sum_{i=1}^{N} dr_{i} = R$) and similarly discretize time coordinate $dt_{n}$ ($\sum_{n=1}^{N^{'}} dt_{n} = t$). Intrinsic quantities such as internal temperature and conductivity should be defined at the spatial zone centres, and the extrinsic quantities such as luminosity at the spatial zone boundaries. The time steps and the spatial zones should be very thin due to the ultra-dense and highly sensitive conductive nature of neutron star layers. To facilitate the numerical simulations further, one can solve the eqs. \ref{eq5} and \ref{eq6} separately by subdividing the star into an outer thin heat blanketing envelope ($\sim$ $100\,m$), which is composed of either heavy or light elements depending on the model used and acts as a thermal insulator for rest of the star. In our numerical conventions, we adopted the backward Euler method with the stated initial and boundary conditions using the surface-boundary temperature relations ($T_{s} - T_{b}$) of the carbon-iron (C-Fe mixture) composed envelope \citep{10.1093/mnras/stw751,BEZNOGOV20211}. The superfluid effects are not considered in the present study. We followed the similar mechanism for cooling simulations as presented by Gnedin {\it et al.}, and the keen readers are advised to track the listed references and the references therein for more detailed technical particularity \citep{1991ApJS...75..449V, 1997A&A...323..415P, 10.1046/j.1365-8711.2001.04359.x}.
\begin{figure}
\centering
\includegraphics[width=1\columnwidth]{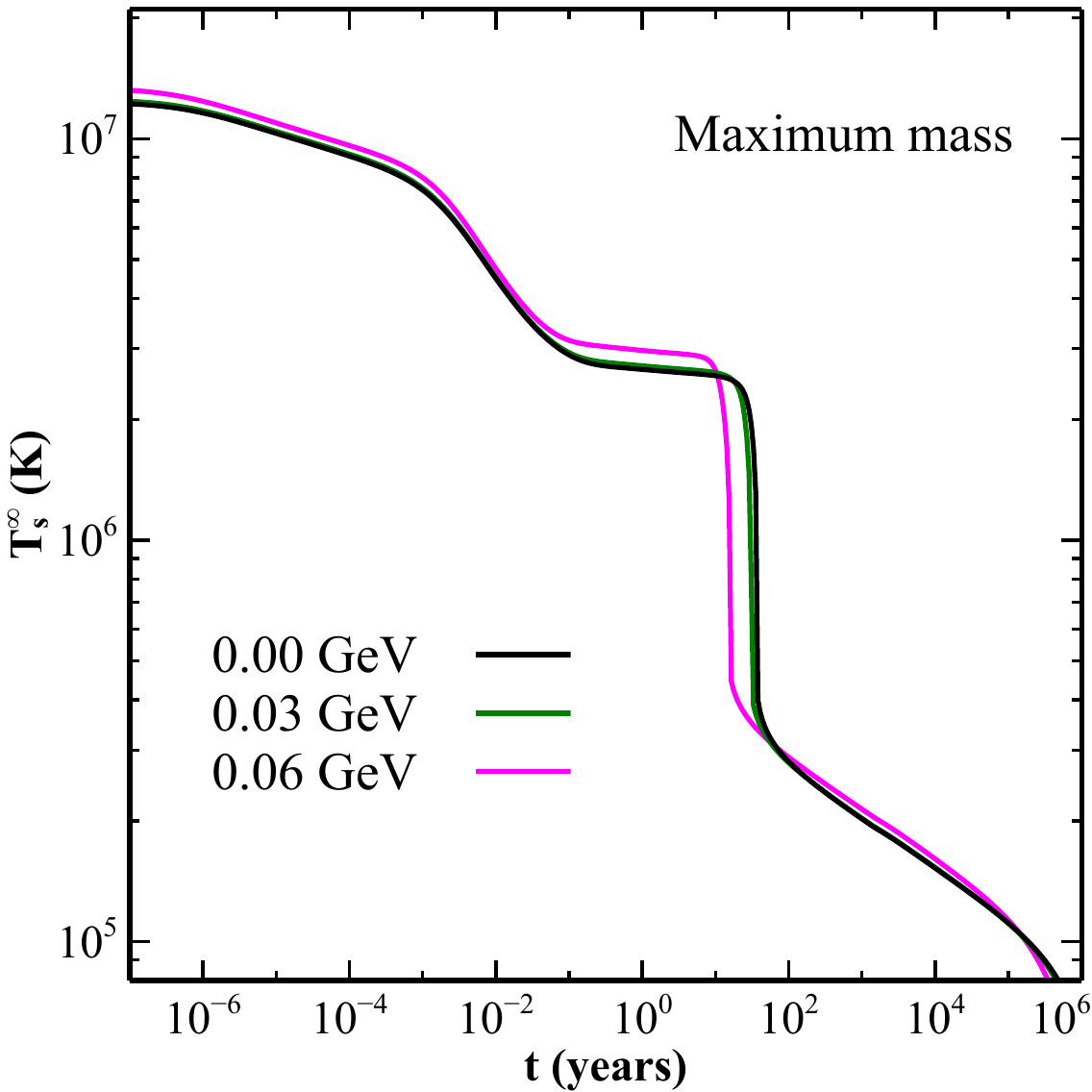}
\caption{Redshifted surface temperature profile of the maximum mass star versus age. The curves for the dark matter admixed maximum mass star profile with $k_{f}^{\rm DM} = 0.00, 0.03\, \& \,0.06$ are shown. The maximum mass of the star decreases with an increase in the dark matter fermi momentum, but the calculated mass for all the cases is greater than $1.4 M_{\odot}$ and so the fast cooling scenario is followed in all the cases.}
\label{fig9}
\end{figure}

\begin{figure*}
\centering
\includegraphics[width=2\columnwidth]{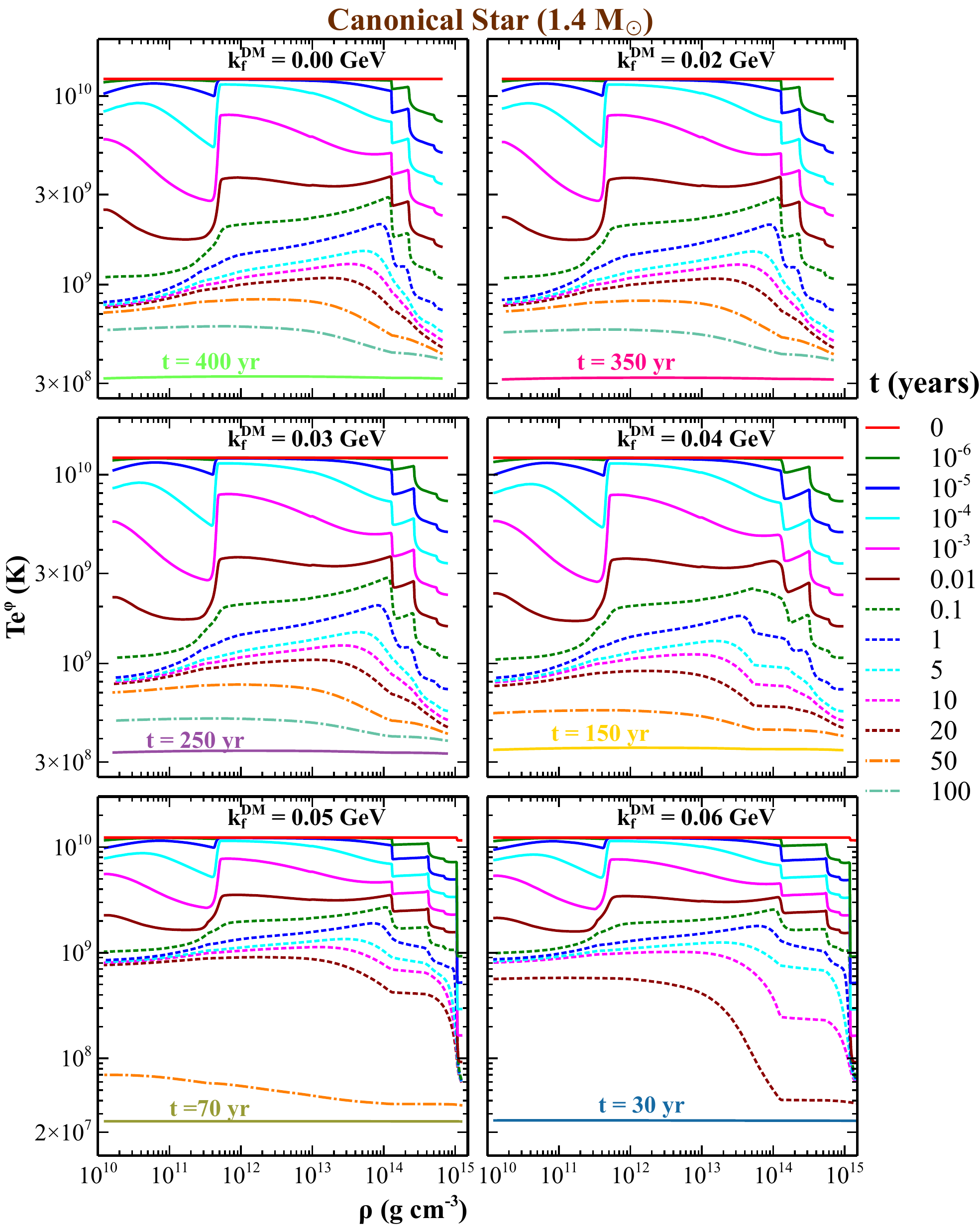}
\caption{Redshifted internal temperature of a dark matter admixed canonical star calculated with the IOPB-I parameter equation of states as a function of density. The dark matter momentum $k_{f}^{\rm DM}$ is varied from $0.00$ GeV to $0.06$ GeV. Each curve corresponds to a specific stellar age and the time required for the thermal relaxation of each segment is mentioned at the bottom. The stellar age key for the curves is presented at the right hand side of the graph.}
\label{fig10}
\end{figure*}

\begin{figure*}
\centering
\includegraphics[width=2\columnwidth]{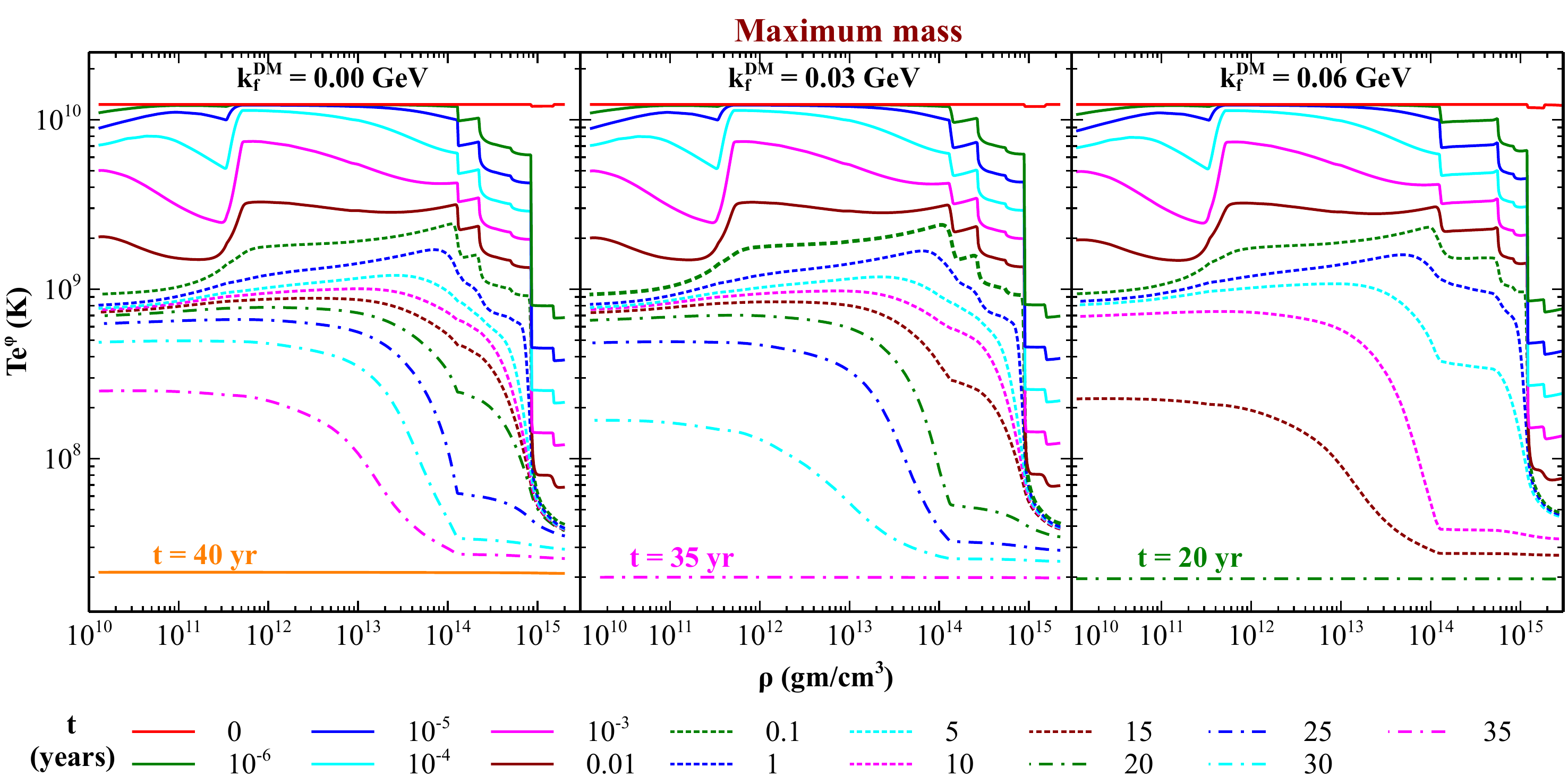}
\caption{Temperature profiles for the interior of a dark matter admixed maximum mass star calculated with the IOPB-I parameter equation of states. The red-shiftted temperature profile curves are illustrated for three different dark matter momentum cases i.e. $k_{f}^{\rm DM} = 0.00,\, 0.03\, \& \,0.06$ GeV. The stellar age key is provided at the bottom and some of the initial time epochs are chosen similar to that of the case for canonical star plot.}
\label{fig11}
\end{figure*}

The results of the numerical simulations for the effective surface temperature ($T_{s}^{\infty}$) as s function of star's age and the redshifted interior temperature ($\Tilde{T} \equiv Te^{\phi}$) as a function of star's density for various stellar age profiles are exemplified in figs. \ref{fig8} - \ref{fig11}. The results are shown for both canonical and maximum mass stars with the different dark matter occupied segments as discussed in the former sections. We also confront the theoretically obtained cooling curves with the thermal emissions of observational pulsar data.
A detailed spectral analysis of several thermally emitting observed pulsars along with the non-thermal originated neutron stars is done by Potekhin {\it et al.} in \citep{10.1093/mnras/staa1871}. The authors in the same reference also discuss the effects of the magnetic field, superfluidity and outer heat blanketing envelope on the extracted values of age and thermal luminosities through various techniques from the observationally recorded data by X-ray laboratories. The cooling era of a neutron star can be recognized in three distinct stages: initial thermal relaxation, neutrino and photon cooling stages \citep{1994ApJ...425..802L}. Initially, the crust segment of the star remains thermally decoupled from the core and the cooling curve reflects the thermal emission dynamics of the crust in the initial years, which is considerably affected by the chemical composition of the envelope \citep{doi:10.1146/annurev.nucl.56.080805.140600}. Thus, the initial cooling stage of both the canonical and maximum mass dark matter admixed star is independent of the EOS or core composition and demonstrates similar cooling pattern for various considered values of dark matter momenta. The neutrino cooling era lasts up to $10^{4}$ - $10^{5}$ years and mainly influenced by the neutrino emissions from the stellar core. The final cooling stage is, however, referred as the photon emissions from the stellar surface and almost insensitive to the core segment of the star, but during final epoch the temperature decreases rapidly due to the overlap of various cooling scenarios and seems to be contingent on the specific heat of the star \citep{1998ASIC..515..539P}. The effects of dark matter admixture on the photon cooling stage of a maximum mass star turn out to be minimal and the cooling curves of the canonical star tend to approach the same region too after $10^{5}$ years of age. This duration of the cooling era also seems to be independent of the EOS of stellar matter obtained through mean-field formalism for various dark matter admixed stars. We reviewed that the cooling rate of the neutron star depends on the compositional distribution of the stellar core and the EOS of the model used to describe the stellar structure rather than the overall mass of the star. Mathematically, the overall luminosity and the integrated heat capacity of the star depend on the mass of the star in the same manner and therefore the effect of the total mass in the energy balance equation cancels out and drives the effective surface temperature insensitive to mass. It is quite straightforward from fig. \ref{fig8} that even if we fix the mass of the star at some constant value and vary only the internal configuration and compositional structure, the cooling rate undergoes thorough the transition from slow cooling to fast cooling scenario. The rapid decrease in the surface temperature of canonical star with dark matter Fermi momentum participation greater than $0.04$ GeV during the neutrino cooling epoch is due to the excess neutrino production in the stellar core as a result of the activation of direct Urca process channel for such stars. In the case of maximum mass star, the direct Urca process is observed for all the configurations (with or without dark matter component), and so the fast cooling tendency is also followed in all such cases (fig. \ref{fig9}). The canonical and maximum mass neutron star configuration with the larger dark matter component turns out to be cool more rapidly as compared to smaller or no dark matter admixed star. The small dip in the apparent surface temperature curve for canonical star with $k_{f}^{\rm DM} < 0.05$ GeV around hundred years of stellar age is the consequence of enhanced neutrino emissivity due to the modified Urca process and the activation of proton branch for the same. The effect of the modified Urca process can also be noticed in the fast cooling scenarios ($k_{f}^{\rm DM} > 0.04$ GeV) around the same stellar age with further scaling down of curves around $3\times10^{5}\, K$ apparent surface temperature. The slope change of the cooling curves around $10^{5}$ years stellar age indicates the transition to the photon cooling stage.

The thermal relaxation of the dark matter admixed canonical and maximum mass stars in terms of internal temperature profiles is illustrated in figs. \ref{fig10} and \ref{fig11} respectively. We observe that up to 5 years of stellar age, the inner crust, outer crust, outer core and inner core of the neutron star form almost decoupled type of thermal reservoirs for all the cases of dark matter admixture, which again manifests the independency of initial cooling of a star on the EOS and internal composition. The effect of fast thermal conduction for the case of high dark matter momentum admixed neutron stars is also reflected in these cooling curves. The canonical star with dark matter momentum $0.06$ GeV achieve internal thermal equilibrium in $30$ years whereas the star with $k_{f}^{\rm DM} = 0.02$ GeV took $350$ years for the outer crust and inner core to be in the same thermal phase. Also, for the case of canonical star with $k_{f}^{\rm DM} = 0.05\, \& \,0.06$ GeV, the central core is much much colder than the outer core and the crustal section of the star, which is the manifestation of powerful direct Urca process in the central core of these stars. A similar kind of scenario with much lower core's temperature can also be noticed in all cases of the maximum mass star profiles. The thermal conduction of a neutron star is not strong enough to overcome the effects of Urca processes and unable to establish a throughout thermal relaxation for several years. The transmission of cooling waves in case of slow and fast cooling scenarios impacts the relaxation mechanism of the crust in a drastic manner. As we mentioned previously too (fig. \ref{fig8}) that a canonical star with $k_{f}^{\rm DM} = 0.05\, \& \,0.06$ GeV follow fast cooling scenario while the other admixed stars cools down slowly. We here noticed that for the fast cooling scenario the temperature gradient of the crustal part is almost eroded after one-tenth of a year, while in the case of slow cooling scenario it took more than $10$ years for the crust temperature profile to dissipate gradient. A huge abrupt drop of surface (crustal) temperature is also observed for all the fast cooling scenarios before attaining internal equilibrium while in the case of slow cooling structure the equilibrium is attained through a consistent reduction.  The effect of neutron dripping on the outer and inner crust equilibrium almost vanishes after $\sim 1$ year in both the scenarios, which means that the dominance of plasmon decay in the neutrino emissivity vanishes and the next epoch will be controlled by the Bremsstrahlung processes.

Overall, despite the large temperature gradient between the crust and core of high dark matter segment canonical and maximum mass stars, the internal thermal relaxation is achieved quicker in these cases. In the case of pure canonical star with IOPB-I parameter set it took $400$ years to attain the thermal relaxation while for the dark matter admixed canonical star with $k_{f}^{\rm DM} = 0.06$ GeV, the same was achieved only in $30$ years of stellar age. Similarly, for the case of maximum mass star the relaxation time of the dark matter admixed star is half as compared to the one without any dark matter admixture. We also realized that the thermal relaxation of a heavier star is much quicker than the lighter one with the same chemical and compositional structure. A pure canonical star with IOPB-I set is thermally relaxed in 400 years while a pure maximum mass star with the same parameter set ($2.130\, M_{\odot}$) instates the equal temperature for crust and core only in 40 years. Similarly, there is a vast difference between the thermal relaxation time period of a canonical and maximum mass dark matter admixed star of $k_{f}^{\rm DM} = 0.03$ GeV. We, therefore, deduce that, however, the surface temperature of a star is almost independent its mass whereas the internal thermal relaxation contradicts this fact and seems to be strongly influenced by the gravitational mass of the star. The thermal relaxation time of a star also depends considerably on the magnetic field and superfluid effects through specific heat and thermal conductivity, which are not adopted in the present simulations. An analytical approach to establish the mathematical estimates between the relaxation time, crust thickness, specific heat, thermal conductivity and radius of the star is also explored in a number of papers \citep{1994ApJ...425..802L, 10.1046/j.1365-8711.2001.04359.x} and it has been emphasized that the crust and core have distinct relaxation times and mechanism, which can also be verified through our present simulations. 
\section{Conclusions}
We have incorporated the neutralino as a dark matter candidate to derive the EOS for a neutron star using RMF formalism and investigated the indirect effects of fermionic dark matter particles on various neutrino emission processes inside a stellar matter. The threshold density for the activation of the direct Urca process inside a canonical star with different compositional structures has been explored. The stellar mass threshold for the direct Urca process to operate is significantly lowered due to the presence of dark matter admixture. The presence of an enhanced neutrino cooling mechanism sends cooling wave from the stellar core to the crust which causes a sudden drop in the surface temperature and mainly responsible for the fast cooling mechanism of the star. We inferred through our theoretical simulations that these powerful neutrino emitting reactions are greatly influenced from the dark matter segment trapping dynamics and configuration. The two other basic physical ingredients, specific heat and thermal conductivity, required for the numerical modeling of cooling dynamics of a neutron star have also been explored. The specific heat of the crustal segment of the star appears to be independent of the dark matter inclusion while the heat capacity of the core is interrupted and further enhanced by the presence of muons along with electrons. The heat conduction in the crust is also less pronounced due to the presence of dark matter candidate, however, it sight to be sensitive to the shape and size of nuclei in the outer crust and the temperature dependency of the thermal conductivity in the crustal segment has also been studied, which increases up to $10^{5}$ orders of magnitude as the temperature is lowered. The thermal relaxation of a canonical star with different admixtures of dark matter momentum is also sensitive to the heat capacity and thermal conduction of the stellar crust and core part, which is not explored separately in the present calculations but the indirect effects on the fast cooling scenarios are manifested in the similar way as depicted by Lattimer {\it et al.} \citep{1994ApJ...425..802L}. Some of the observational limits for the surface temperature of the canonical star also seem to be consistent with the current theoretical simulations. The variation in the dark matter admixtures of the canonical star vastly influences the surface temperature decrement with the stellar age, which solidifies the fact that the cooling mechanism depends on the internal composition and EOS too, rather than solely on the gravitational mass. We also observed that the internal thermal relaxation time period of the star is very sensitive to the percentage of dark matter segment and the theoretical consideration of greater dark matter percentage enhances both the surface temperature and the internal relaxation of the star drastically.

Apart from the present scenario, there are a number of opportunities in which the present study can be extended. To begin with, we did not consider the reheating mechanism of old neutron stars due to the accretion of dark matter and heat dissipation due to the very high rotational frequency of the stars. The energy released due to the annihilation of dark matter can expedite the reheating dynamics which can also be responsible for the core-collapse and gravitational bursts. The interesting possibility of the presence of strange hyperons, phase transition to quark matter and kaons had also not been considered. The cooper pairing, superfluidity and high magnetic field of the neutron star can also be included in the simulations for a more practical picture of cooling dynamics. Detailed exploration of all these possibilities along with the dark matter inclusion is beyond the scope of present work and may be investigated in future projects.  
\section*{Data Availability}
This manuscript has no associated data and data sharing is not applicable to this article as no data sets were generated during the current study.
\bibliography{thermevo}

\begin{thebibliography}{}
\makeatletter
\relax
\def\mn@urlcharsother{\let\do\@makeother \do\$\do\&\do\#\do\^\do\_\do\%\do\~}
\def\mn@doi{\begingroup\mn@urlcharsother \@ifnextchar [ {\mn@doi@}
  {\mn@doi@[]}}
\def\mn@doi@[#1]#2{\def\@tempa{#1}\ifx\@tempa\@empty \href
  {http://dx.doi.org/#2} {doi:#2}\else \href {http://dx.doi.org/#2} {#1}\fi
  \endgroup}
\def\mn@eprint#1#2{\mn@eprint@#1:#2::\@nil}
\def\mn@eprint@arXiv#1{\href {http://arxiv.org/abs/#1} {{\tt arXiv:#1}}}
\def\mn@eprint@dblp#1{\href {http://dblp.uni-trier.de/rec/bibtex/#1.xml}
  {dblp:#1}}
\def\mn@eprint@#1:#2:#3:#4\@nil{\def\@tempa {#1}\def\@tempb {#2}\def\@tempc
  {#3}\ifx \@tempc \@empty \let \@tempc \@tempb \let \@tempb \@tempa \fi \ifx
  \@tempb \@empty \def\@tempb {arXiv}\fi \@ifundefined
  {mn@eprint@\@tempb}{\@tempb:\@tempc}{\expandafter \expandafter \csname
  mn@eprint@\@tempb\endcsname \expandafter{\@tempc}}}

\bibitem[\protect\citeauthoryear{Abbott et~al.,}{Abbott
  et~al.}{2017}]{PhysRevLett.119.161101}
Abbott B.~P.,  et~al., 2017, \mn@doi [Phys. Rev. Lett.]
  {10.1103/PhysRevLett.119.161101}, 119, 161101

\bibitem[\protect\citeauthoryear{{Abbott} et~al.,}{{Abbott}
  et~al.}{2020}]{Abott_2020}
{Abbott} R.,  et~al., 2020, \mn@doi [\apjl] {10.3847/2041-8213/ab960f}, \href
  {https://ui.adsabs.harvard.edu/abs/2020ApJ...896L..44A} {896, L44}

\bibitem[\protect\citeauthoryear{Alvarez-Ruso, Ledwig, Martin~Camalich  \&
  Vicente-Vacas}{Alvarez-Ruso et~al.}{2013}]{PhysRevD.88.054507}
Alvarez-Ruso L.,  Ledwig T.,  Martin~Camalich J.,   Vicente-Vacas M.~J.,  2013,
  \mn@doi [Phys. Rev. D] {10.1103/PhysRevD.88.054507}, 88, 054507

\bibitem[\protect\citeauthoryear{Baer, Barger, Huang, Mickelson,
  Padeffke-Kirkland  \& Tata}{Baer et~al.}{2015}]{PhysRevD.91.075005}
Baer H.,  Barger V.,  Huang P.,  Mickelson D.,  Padeffke-Kirkland M.,   Tata
  X.,  2015, \mn@doi [Phys. Rev. D] {10.1103/PhysRevD.91.075005}, 91, 075005

\bibitem[\protect\citeauthoryear{Baiko, Potekhin  \& Yakovlev}{Baiko
  et~al.}{2001a}]{PhysRevE.64.057402}
Baiko D.~A.,  Potekhin A.~Y.,   Yakovlev D.~G.,  2001a, \mn@doi [Phys. Rev. E]
  {10.1103/PhysRevE.64.057402}, 64, 057402

\bibitem[\protect\citeauthoryear{{Baiko}, {Haensel}  \& {Yakovlev}}{{Baiko}
  et~al.}{2001b}]{2001A&A...374..151B}
{Baiko} D.~A.,  {Haensel} P.,   {Yakovlev} D.~G.,  2001b, \mn@doi [\aap]
  {10.1051/0004-6361:20010621}, \href
  {https://ui.adsabs.harvard.edu/abs/2001A&A...374..151B} {374, 151}

\bibitem[\protect\citeauthoryear{Bertone, Hooper  \& Silk}{Bertone
  et~al.}{2005}]{Bertone_2005}
Bertone G.,  Hooper D.,   Silk J.,  2005, \mn@doi [Physics Reports]
  {10.1016/j.physrep.2004.08.031}, 405, 279–390

\bibitem[\protect\citeauthoryear{Beznogov, Potekhin  \& Yakovlev}{Beznogov
  et~al.}{2016}]{10.1093/mnras/stw751}
Beznogov M.~V.,  Potekhin A.~Y.,   Yakovlev D.~G.,  2016, \mn@doi [Monthly
  Notices of the Royal Astronomical Society] {10.1093/mnras/stw751}, 459, 1569

\bibitem[\protect\citeauthoryear{Beznogov, Potekhin  \& Yakovlev}{Beznogov
  et~al.}{2021}]{BEZNOGOV20211}
Beznogov M.,  Potekhin A.,   Yakovlev D.,  2021, \mn@doi [Physics Reports]
  {https://doi.org/10.1016/j.physrep.2021.03.004}, 919, 1

\bibitem[\protect\citeauthoryear{Bilenky \& Hošek}{Bilenky \&
  Hošek}{1982}]{BILENKY198273}
Bilenky S.,  Hošek J.,  1982, \mn@doi [Physics Reports]
  {https://doi.org/10.1016/0370-1573(82)90016-3}, 90, 73

\bibitem[\protect\citeauthoryear{Bisnovatyi-Kogan \& Romanova}{Bisnovatyi-Kogan
  \& Romanova}{1982}]{Bisnovatyi-Kogan1982}
Bisnovatyi-Kogan G.~S.,  Romanova M.~M.,  1982, Soviet Physics - JETP, 56, 243

\bibitem[\protect\citeauthoryear{{Bludman} \& {Ruderman}}{{Bludman} \&
  {Ruderman}}{1975}]{1975ApJ...195L..19B}
{Bludman} S.~A.,  {Ruderman} M.~A.,  1975, \mn@doi [\apjl] {10.1086/181699},
  \href {https://ui.adsabs.harvard.edu/abs/1975ApJ...195L..19B} {195, L19}

\bibitem[\protect\citeauthoryear{Boguta}{Boguta}{1981}]{BOGUTA1981255}
Boguta J.,  1981, \mn@doi [Physics Letters B]
  {https://doi.org/10.1016/0370-2693(81)90529-3}, 106, 255

\bibitem[\protect\citeauthoryear{Bowyer, Byram, Chubb  \& Friedman}{Bowyer
  et~al.}{1964}]{doi:10.1126/science.146.3646.912}
Bowyer S.,  Byram E.~T.,  Chubb T.~A.,   Friedman H.,  1964, \mn@doi [Science]
  {10.1126/science.146.3646.912}, 146, 912

\bibitem[\protect\citeauthoryear{Braaten}{Braaten}{1991}]{PhysRevLett.66.1655}
Braaten E.,  1991, \mn@doi [Phys. Rev. Lett.] {10.1103/PhysRevLett.66.1655},
  66, 1655

\bibitem[\protect\citeauthoryear{{Burrows} \& {Lattimer}}{{Burrows} \&
  {Lattimer}}{1986}]{1986ApJ...307..178B}
{Burrows} A.,  {Lattimer} J.~M.,  1986, \mn@doi [\apj] {10.1086/164405}, \href
  {https://ui.adsabs.harvard.edu/abs/1986ApJ...307..178B} {307, 178}

\bibitem[\protect\citeauthoryear{Carbone \& Schwenk}{Carbone \&
  Schwenk}{2019}]{PhysRevC.100.025805}
Carbone A.,  Schwenk A.,  2019, \mn@doi [Phys. Rev. C]
  {10.1103/PhysRevC.100.025805}, 100, 025805

\bibitem[\protect\citeauthoryear{Carr}{Carr}{1961}]{PhysRev.122.1437}
Carr W.~J.,  1961, \mn@doi [Phys. Rev.] {10.1103/PhysRev.122.1437}, 122, 1437

\bibitem[\protect\citeauthoryear{Chamel \& Haensel}{Chamel \&
  Haensel}{2008}]{Chamel2008}
Chamel N.,  Haensel P.,  2008, \mn@doi [Living Reviews in Relativity]
  {10.12942/lrr-2008-10}, 11, 10

\bibitem[\protect\citeauthoryear{Chiu \& Morrison}{Chiu \&
  Morrison}{1960}]{PhysRevLett.5.573}
Chiu H.-Y.,  Morrison P.,  1960, \mn@doi [Phys. Rev. Lett.]
  {10.1103/PhysRevLett.5.573}, 5, 573

\bibitem[\protect\citeauthoryear{Chiu \& Salpeter}{Chiu \&
  Salpeter}{1964}]{PhysRevLett.12.413}
Chiu H.-Y.,  Salpeter E.~E.,  1964, \mn@doi [Phys. Rev. Lett.]
  {10.1103/PhysRevLett.12.413}, 12, 413

\bibitem[\protect\citeauthoryear{Cline, Scott, Kainulainen  \& Weniger}{Cline
  et~al.}{2013}]{PhysRevD.88.055025}
Cline J.~M.,  Scott P.,  Kainulainen K.,   Weniger C.,  2013, \mn@doi [Phys.
  Rev. D] {10.1103/PhysRevD.88.055025}, 88, 055025

\bibitem[\protect\citeauthoryear{Collins \& Perry}{Collins \&
  Perry}{1975}]{PhysRevLett.34.1353}
Collins J.~C.,  Perry M.~J.,  1975, \mn@doi [Phys. Rev. Lett.]
  {10.1103/PhysRevLett.34.1353}, 34, 1353

\bibitem[\protect\citeauthoryear{Das, Malik  \& Nayak}{Das
  et~al.}{2019}]{PhysRevD.99.043016}
Das A.,  Malik T.,   Nayak A.~C.,  2019, \mn@doi [Phys. Rev. D]
  {10.1103/PhysRevD.99.043016}, 99, 043016

\bibitem[\protect\citeauthoryear{Das, Kumar, Kumar, Biswal, Nakatsukasa, Li  \&
  Patra}{Das et~al.}{2020}]{10.1093/mnras/staa1435}
Das H.~C.,  Kumar A.,  Kumar B.,  Biswal S.~K.,  Nakatsukasa T.,  Li A.,
  Patra S.~K.,  2020, \mn@doi [Monthly Notices of the Royal Astronomical
  Society] {10.1093/mnras/staa1435}, 495, 4893

\bibitem[\protect\citeauthoryear{Das, Kumar, Biswal  \& Patra}{Das
  et~al.}{2021a}]{PhysRevD.104.123006}
Das H.~C.,  Kumar A.,  Biswal S.~K.,   Patra S.~K.,  2021a, \mn@doi [Phys. Rev.
  D] {10.1103/PhysRevD.104.123006}, 104, 123006

\bibitem[\protect\citeauthoryear{Das, Kumar  \& Patra}{Das
  et~al.}{2021b}]{10.1093/mnras/stab2387}
Das H.~C.,  Kumar A.,   Patra S.~K.,  2021b, \mn@doi [Monthly Notices of the
  Royal Astronomical Society] {10.1093/mnras/stab2387}, 507, 4053

\bibitem[\protect\citeauthoryear{Das, Kumar, Kumar, Biswal  \& Patra}{Das
  et~al.}{2021c}]{Das_2021}
Das H.,  Kumar A.,  Kumar B.,  Biswal S.,   Patra S.,  2021c, \mn@doi [Journal
  of Cosmology and Astroparticle Physics] {10.1088/1475-7516/2021/01/007},
  2021, 007

\bibitem[\protect\citeauthoryear{Dhiman, Kumar  \& Agrawal}{Dhiman
  et~al.}{2007}]{PhysRevC.76.045801}
Dhiman S.~K.,  Kumar R.,   Agrawal B.~K.,  2007, \mn@doi [Phys. Rev. C]
  {10.1103/PhysRevC.76.045801}, 76, 045801

\bibitem[\protect\citeauthoryear{Duffy \& Bibber}{Duffy \&
  Bibber}{2009}]{Leanne_2009}
Duffy L.~D.,  Bibber K.~v.,  2009, \mn@doi [New Journal of Physics]
  {10.1088/1367-2630/11/10/105008}, 11, 105008

\bibitem[\protect\citeauthoryear{Dutra, Lourenço, Sá~Martins, Delfino, Stone
  \& Stevenson}{Dutra et~al.}{2012}]{Dutra_2012}
Dutra M.,  Lourenço O.,  Sá~Martins J.~S.,  Delfino A.,  Stone J.~R.,
  Stevenson P.~D.,  2012, \mn@doi [Physical Review C]
  {10.1103/physrevc.85.035201}, 85

\bibitem[\protect\citeauthoryear{Dutra et~al.,}{Dutra
  et~al.}{2014}]{PhysRevC.90.055203}
Dutra M.,  et~al., 2014, \mn@doi [Phys. Rev. C] {10.1103/PhysRevC.90.055203},
  90, 055203

\bibitem[\protect\citeauthoryear{Dutra, Louren\ifmmode~\mbox{\c{c}}\else
  \c{c}\fi{}o  \& Menezes}{Dutra et~al.}{2016}]{PhysRevC.93.025806}
Dutra M.,  Louren\ifmmode~\mbox{\c{c}}\else \c{c}\fi{}o O.,   Menezes D.~P.,
  2016, \mn@doi [Phys. Rev. C] {10.1103/PhysRevC.93.025806}, 93, 025806

\bibitem[\protect\citeauthoryear{Ellis, Hektor, Hütsi, Kannike, Marzola,
  Raidal  \& Vaskonen}{Ellis et~al.}{2018}]{ELLIS2018607}
Ellis J.,  Hektor A.,  Hütsi G.,  Kannike K.,  Marzola L.,  Raidal M.,
  Vaskonen V.,  2018, \mn@doi [Physics Letters B]
  {https://doi.org/10.1016/j.physletb.2018.04.048}, 781, 607

\bibitem[\protect\citeauthoryear{Entem, Kaiser, Machleidt  \& Nosyk}{Entem
  et~al.}{2015}]{PhysRevC.91.014002}
Entem D.~R.,  Kaiser N.,  Machleidt R.,   Nosyk Y.,  2015, \mn@doi [Phys. Rev.
  C] {10.1103/PhysRevC.91.014002}, 91, 014002

\bibitem[\protect\citeauthoryear{Epelbaum, Hammer  \& Mei\ss{}ner}{Epelbaum
  et~al.}{2009}]{RevModPhys.81.1773}
Epelbaum E.,  Hammer H.-W.,   Mei\ss{}ner U.-G.,  2009, \mn@doi [Rev. Mod.
  Phys.] {10.1103/RevModPhys.81.1773}, 81, 1773

\bibitem[\protect\citeauthoryear{{Fan}, {Yang}  \& {Chang}}{{Fan}
  et~al.}{2012}]{2012arXiv1204.2564F}
{Fan} Y.-z.,  {Yang} R.-z.,   {Chang} J.,  2012, arXiv e-prints, \href
  {https://ui.adsabs.harvard.edu/abs/2012arXiv1204.2564F} {p. arXiv:1204.2564}

\bibitem[\protect\citeauthoryear{Fattoyev, Horowitz, Piekarewicz  \&
  Shen}{Fattoyev et~al.}{2010}]{PhysRevC.82.055803}
Fattoyev F.~J.,  Horowitz C.~J.,  Piekarewicz J.,   Shen G.,  2010, \mn@doi
  [Phys. Rev. C] {10.1103/PhysRevC.82.055803}, 82, 055803

\bibitem[\protect\citeauthoryear{{Flowers}}{{Flowers}}{1973}]{1973ApJ...180..911F}
{Flowers} E.,  1973, \mn@doi [\apj] {10.1086/152017}, \href
  {https://ui.adsabs.harvard.edu/abs/1973ApJ...180..911F} {180, 911}

\bibitem[\protect\citeauthoryear{{Flowers} \& {Itoh}}{{Flowers} \&
  {Itoh}}{1976}]{1976ApJ...206..218F}
{Flowers} E.,  {Itoh} N.,  1976, \mn@doi [\apj] {10.1086/154375}, \href
  {https://ui.adsabs.harvard.edu/abs/1976ApJ...206..218F} {206, 218}

\bibitem[\protect\citeauthoryear{{Flowers} \& {Sutherland}}{{Flowers} \&
  {Sutherland}}{1977}]{1977Ap&SS..48..159F}
{Flowers} E.~G.,  {Sutherland} P.~G.,  1977, \mn@doi [\apss]
  {10.1007/BF00643047}, \href
  {https://ui.adsabs.harvard.edu/abs/1977Ap&SS..48..159F} {48, 159}

\bibitem[\protect\citeauthoryear{{Friman} \& {Maxwell}}{{Friman} \&
  {Maxwell}}{1979}]{1979ApJ...232..541F}
{Friman} B.~L.,  {Maxwell} O.~V.,  1979, \mn@doi [\apj] {10.1086/157313}, \href
  {https://ui.adsabs.harvard.edu/abs/1979ApJ...232..541F} {232, 541}

\bibitem[\protect\citeauthoryear{{Fujimoto}, {Hanawa}, {Iben}  \&
  {Richardson}}{{Fujimoto} et~al.}{1984}]{1984ApJ...278..813F}
{Fujimoto} M.~Y.,  {Hanawa} T.,  {Iben} I. J.,   {Richardson} M.~B.,  1984,
  \mn@doi [\apj] {10.1086/161851}, \href
  {https://ui.adsabs.harvard.edu/abs/1984ApJ...278..813F} {278, 813}

\bibitem[\protect\citeauthoryear{Furnstahl, Serot  \& Tang}{Furnstahl
  et~al.}{1997}]{FURNSTAHL1997441}
Furnstahl R.,  Serot B.~D.,   Tang H.-B.,  1997, \mn@doi [Nuclear Physics A]
  {https://doi.org/10.1016/S0375-9474(96)00472-1}, 615, 441

\bibitem[\protect\citeauthoryear{Gambhir, Ring  \& Thimet}{Gambhir
  et~al.}{1990}]{GAMBHIR1990132}
Gambhir Y.,  Ring P.,   Thimet A.,  1990, \mn@doi [Annals of Physics]
  {https://doi.org/10.1016/0003-4916(90)90330-Q}, 198, 132

\bibitem[\protect\citeauthoryear{Gamow \& Schoenberg}{Gamow \&
  Schoenberg}{1941}]{PhysRev.59.539}
Gamow G.,  Schoenberg M.,  1941, \mn@doi [Phys. Rev.] {10.1103/PhysRev.59.539},
  59, 539

\bibitem[\protect\citeauthoryear{Giacconi, Gursky, Paolini  \& Rossi}{Giacconi
  et~al.}{1962}]{PhysRevLett.9.439}
Giacconi R.,  Gursky H.,  Paolini F.~R.,   Rossi B.~B.,  1962, \mn@doi [Phys.
  Rev. Lett.] {10.1103/PhysRevLett.9.439}, 9, 439

\bibitem[\protect\citeauthoryear{{Giacconi} et~al.,}{{Giacconi}
  et~al.}{1979}]{1979ApJ...234L...1G}
{Giacconi} R.,  et~al., 1979, \mn@doi [\apjl] {10.1086/183099}, \href
  {https://ui.adsabs.harvard.edu/abs/1979ApJ...234L...1G} {234, L1}

\bibitem[\protect\citeauthoryear{{Glen} \& {Sutherland}}{{Glen} \&
  {Sutherland}}{1980}]{1980ApJ...239..671G}
{Glen} G.,  {Sutherland} P.,  1980, \mn@doi [\apj] {10.1086/158154}, \href
  {https://ui.adsabs.harvard.edu/abs/1980ApJ...239..671G} {239, 671}

\bibitem[\protect\citeauthoryear{{Glendenning}}{{Glendenning}}{1985}]{1985ApJ...293..470G}
{Glendenning} N.~K.,  1985, \mn@doi [\apj] {10.1086/163253}, \href
  {https://ui.adsabs.harvard.edu/abs/1985ApJ...293..470G} {293, 470}

\bibitem[\protect\citeauthoryear{Glendenning \& Moszkowski}{Glendenning \&
  Moszkowski}{1991}]{PhysRevLett.67.2414}
Glendenning N.~K.,  Moszkowski S.~A.,  1991, \mn@doi [Phys. Rev. Lett.]
  {10.1103/PhysRevLett.67.2414}, 67, 2414

\bibitem[\protect\citeauthoryear{{Gnedin} \& {Yakovlev}}{{Gnedin} \&
  {Yakovlev}}{1995a}]{1995NuPhA.582..697G}
{Gnedin} O.~Y.,  {Yakovlev} D.~G.,  1995a, \mn@doi [\nphysa]
  {10.1016/0375-9474(94)00503-F}, \href
  {https://ui.adsabs.harvard.edu/abs/1995NuPhA.582..697G} {582, 697}

\bibitem[\protect\citeauthoryear{Gnedin \& Yakovlev}{Gnedin \&
  Yakovlev}{1995b}]{GNEDIN1995697}
Gnedin O.,  Yakovlev D.,  1995b, \mn@doi [Nuclear Physics A]
  {https://doi.org/10.1016/0375-9474(94)00503-F}, 582, 697

\bibitem[\protect\citeauthoryear{Gnedin, Yakovlev  \& Potekhin}{Gnedin
  et~al.}{2001}]{10.1046/j.1365-8711.2001.04359.x}
Gnedin O.~Y.,  Yakovlev D.~G.,   Potekhin A.~Y.,  2001, \mn@doi [Monthly
  Notices of the Royal Astronomical Society]
  {10.1046/j.1365-8711.2001.04359.x}, 324, 725

\bibitem[\protect\citeauthoryear{{Gold}}{{Gold}}{1968}]{1968Natur.218..731G}
{Gold} T.,  1968, \mn@doi [\nat] {10.1038/218731a0}, \href
  {https://ui.adsabs.harvard.edu/abs/1968Natur.218..731G} {218, 731}

\bibitem[\protect\citeauthoryear{Goldman \& Nussinov}{Goldman \&
  Nussinov}{1989}]{PhysRevD.40.3221}
Goldman I.,  Nussinov S.,  1989, \mn@doi [Phys. Rev. D]
  {10.1103/PhysRevD.40.3221}, 40, 3221

\bibitem[\protect\citeauthoryear{Greif, Hebeler, Lattimer, Pethick  \&
  Schwenk}{Greif et~al.}{2020}]{Greif_2020}
Greif S.~K.,  Hebeler K.,  Lattimer J.~M.,  Pethick C.~J.,   Schwenk A.,  2020,
  \mn@doi [The Astrophysical Journal] {10.3847/1538-4357/abaf55}, 901, 155

\bibitem[\protect\citeauthoryear{{Haensel}, {Kaminker}  \&
  {Yakovlev}}{{Haensel} et~al.}{1996}]{1996A&A...314..328H}
{Haensel} P.,  {Kaminker} A.~D.,   {Yakovlev} D.~G.,  1996, \aap, \href
  {https://ui.adsabs.harvard.edu/abs/1996A&A...314..328H} {314, 328}

\bibitem[\protect\citeauthoryear{Ho \& Heinke}{Ho \& Heinke}{2009}]{Ho_2009}
Ho W. C.~G.,  Heinke C.~O.,  2009, \mn@doi [Nature] {10.1038/nature08525}, 462,
  71

\bibitem[\protect\citeauthoryear{Hong, Hsu  \& Sannino}{Hong
  et~al.}{2001}]{Hong}
Hong D.~K.,  Hsu S.~D.,   Sannino F.,  2001, \mn@doi [Physics Letters B]
  {10.1016/s0370-2693(01)00955-8}, 516, 362–366

\bibitem[\protect\citeauthoryear{Horowitz \& Piekarewicz}{Horowitz \&
  Piekarewicz}{2002}]{PhysRevC.66.055803}
Horowitz C.~J.,  Piekarewicz J.,  2002, \mn@doi [Phys. Rev. C]
  {10.1103/PhysRevC.66.055803}, 66, 055803

\bibitem[\protect\citeauthoryear{{Itoh}, {Mutoh}, {Hikita}  \&
  {Kohyama}}{{Itoh} et~al.}{1992}]{1992ApJ...395..622I}
{Itoh} N.,  {Mutoh} H.,  {Hikita} A.,   {Kohyama} Y.,  1992, \mn@doi [\apj]
  {10.1086/171682}, \href
  {https://ui.adsabs.harvard.edu/abs/1992ApJ...395..622I} {395, 622}

\bibitem[\protect\citeauthoryear{Jungman, Kamionkowski  \& Griest}{Jungman
  et~al.}{1996}]{Jungman_1996}
Jungman G.,  Kamionkowski M.,   Griest K.,  1996, \mn@doi [Physics Reports]
  {10.1016/0370-1573(95)00058-5}, 267, 195–373

\bibitem[\protect\citeauthoryear{{Kaminker}, {Pethick}, {Potekhin}, {Thorsson}
  \& {Yakovlev}}{{Kaminker} et~al.}{1999}]{1999A&A...343.1009K}
{Kaminker} A.~D.,  {Pethick} C.~J.,  {Potekhin} A.~Y.,  {Thorsson} V.,
  {Yakovlev} D.~G.,  1999, \aap, \href
  {https://ui.adsabs.harvard.edu/abs/1999A&A...343.1009K} {343, 1009}

\bibitem[\protect\citeauthoryear{Kantor \& Gusakov}{Kantor \&
  Gusakov}{2007}]{10.1111/j.1365-2966.2007.12342.x}
Kantor E.~M.,  Gusakov M.~E.,  2007, \mn@doi [Monthly Notices of the Royal
  Astronomical Society] {10.1111/j.1365-2966.2007.12342.x}, 381, 1702

\bibitem[\protect\citeauthoryear{Kaplan \& Nelson}{Kaplan \&
  Nelson}{1988}]{KAPLAN1988273}
Kaplan D.,  Nelson A.,  1988, \mn@doi [Nuclear Physics A]
  {https://doi.org/10.1016/0375-9474(88)90442-3}, 479, 273

\bibitem[\protect\citeauthoryear{Kouvaris}{Kouvaris}{2008}]{PhysRevD.77.023006}
Kouvaris C.,  2008, \mn@doi [Phys. Rev. D] {10.1103/PhysRevD.77.023006}, 77,
  023006

\bibitem[\protect\citeauthoryear{Kouvaris \& Tinyakov}{Kouvaris \&
  Tinyakov}{2010}]{PhysRevD.82.063531}
Kouvaris C.,  Tinyakov P.,  2010, \mn@doi [Phys. Rev. D]
  {10.1103/PhysRevD.82.063531}, 82, 063531

\bibitem[\protect\citeauthoryear{Kumar, Agrawal  \& Dhiman}{Kumar
  et~al.}{2006}]{PhysRevC.74.034323}
Kumar R.,  Agrawal B.~K.,   Dhiman S.~K.,  2006, \mn@doi [Phys. Rev. C]
  {10.1103/PhysRevC.74.034323}, 74, 034323

\bibitem[\protect\citeauthoryear{Kumar, Singh, Agrawal  \& Patra}{Kumar
  et~al.}{2017}]{KUMAR2017197}
Kumar B.,  Singh S.,  Agrawal B.,   Patra S.,  2017, \mn@doi [Nuclear Physics
  A] {https://doi.org/10.1016/j.nuclphysa.2017.07.001}, 966, 197

\bibitem[\protect\citeauthoryear{Kumar, Patra  \& Agrawal}{Kumar
  et~al.}{2018}]{PhysRevC.97.045806}
Kumar B.,  Patra S.~K.,   Agrawal B.~K.,  2018, \mn@doi [Phys. Rev. C]
  {10.1103/PhysRevC.97.045806}, 97, 045806

\bibitem[\protect\citeauthoryear{Kumar, Das, Biswal, Kumar  \& Patra}{Kumar
  et~al.}{2020}]{Kumar2020}
Kumar A.,  Das H.~C.,  Biswal S.~K.,  Kumar B.,   Patra S.~K.,  2020, \mn@doi
  [The European Physical Journal C] {10.1140/epjc/s10052-020-8353-4}, 80, 775

\bibitem[\protect\citeauthoryear{Kumar, Das, Bhuyan  \& Patra}{Kumar
  et~al.}{2021}]{KUMAR2021122315}
Kumar A.,  Das H.,  Bhuyan M.,   Patra S.,  2021, \mn@doi [Nuclear Physics A]
  {https://doi.org/10.1016/j.nuclphysa.2021.122315}, 1015, 122315

\bibitem[\protect\citeauthoryear{Lalazissis, K\"onig  \& Ring}{Lalazissis
  et~al.}{1997}]{PhysRevC.55.540}
Lalazissis G.~A.,  K\"onig J.,   Ring P.,  1997, \mn@doi [Phys. Rev. C]
  {10.1103/PhysRevC.55.540}, 55, 540

\bibitem[\protect\citeauthoryear{Lattimer}{Lattimer}{2012}]{doi:10.1146/annurev-nucl-102711-095018}
Lattimer J.~M.,  2012, \mn@doi [Annual Review of Nuclear and Particle Science]
  {10.1146/annurev-nucl-102711-095018}, 62, 485

\bibitem[\protect\citeauthoryear{Lattimer, Pethick, Prakash  \&
  Haensel}{Lattimer et~al.}{1991}]{PhysRevLett.66.2701}
Lattimer J.~M.,  Pethick C.~J.,  Prakash M.,   Haensel P.,  1991, \mn@doi
  [Phys. Rev. Lett.] {10.1103/PhysRevLett.66.2701}, 66, 2701

\bibitem[\protect\citeauthoryear{{Lattimer}, {van Riper}, {Prakash}  \&
  {Prakash}}{{Lattimer} et~al.}{1994}]{1994ApJ...425..802L}
{Lattimer} J.~M.,  {van Riper} K.~A.,  {Prakash} M.,   {Prakash} M.,  1994,
  \mn@doi [\apj] {10.1086/174025}, \href
  {https://ui.adsabs.harvard.edu/abs/1994ApJ...425..802L} {425, 802}

\bibitem[\protect\citeauthoryear{Levin \& Ushomirsky}{Levin \&
  Ushomirsky}{2001}]{10.1046/j.1365-8711.2001.04323.x}
Levin Y.,  Ushomirsky G.,  2001, \mn@doi [Monthly Notices of the Royal
  Astronomical Society] {10.1046/j.1365-8711.2001.04323.x}, 324, 917

\bibitem[\protect\citeauthoryear{Li, Mao, Zhuo  \& Greiner}{Li
  et~al.}{1997}]{PhysRevC.56.1570}
Li Z.,  Mao G.,  Zhuo Y.,   Greiner W.,  1997, \mn@doi [Phys. Rev. C]
  {10.1103/PhysRevC.56.1570}, 56, 1570

\bibitem[\protect\citeauthoryear{Li, Huang  \& Xu}{Li et~al.}{2012a}]{LI201270}
Li A.,  Huang F.,   Xu R.-X.,  2012a, \mn@doi [Astroparticle Physics]
  {https://doi.org/10.1016/j.astropartphys.2012.07.006}, 37, 70

\bibitem[\protect\citeauthoryear{Li, Wang  \& Cheng}{Li
  et~al.}{2012b}]{Li_2012}
Li X.,  Wang F.,   Cheng K.,  2012b, \mn@doi [Journal of Cosmology and
  Astroparticle Physics] {10.1088/1475-7516/2012/10/031}, 2012, 031–031

\bibitem[\protect\citeauthoryear{{Liu}, {Guo}, {di Toro}  \& {Greco}}{{Liu}
  et~al.}{2005}]{2005EPJA...25..293L}
{Liu} B.,  {Guo} H.,  {di Toro} M.,   {Greco} V.,  2005, \mn@doi [European
  Physical Journal A] {10.1140/epja/i2005-10095-1}, \href
  {https://ui.adsabs.harvard.edu/abs/2005EPJA...25..293L} {25, 293}

\bibitem[\protect\citeauthoryear{Lourenço, Frederico  \& Dutra}{Lourenço
  et~al.}{2021}]{lourenco2021dark}
Lourenço O.,  Frederico T.,   Dutra M.,  2021, Dark matter component in
  hadronic models with short-range correlations (\mn@eprint {arXiv}
  {2112.07716})

\bibitem[\protect\citeauthoryear{Martin}{Martin}{1998}]{MARTIN_1998}
Martin S.~P.,  1998, \mn@doi [Advanced Series on Directions in High Energy
  Physics (Perspectives On Supersymmetry)] {10.1142/9789812839657_0001}, 18,
  1–98

\bibitem[\protect\citeauthoryear{McGowan, Zane, Cropper, Kennea, Cordova, Ho,
  Sasseen  \& Vestrand}{McGowan et~al.}{2004}]{McGowan_2004}
McGowan K.~E.,  Zane S.,  Cropper M.,  Kennea J.~A.,  Cordova F.~A.,  Ho C.,
  Sasseen T.,   Vestrand W.~T.,  2004, \mn@doi [The Astrophysical Journal]
  {10.1086/379787}, 600, 343

\bibitem[\protect\citeauthoryear{McGowan, Zane, Cropper, Vestrand  \&
  Ho}{McGowan et~al.}{2006}]{McGowan_2006}
McGowan K.~E.,  Zane S.,  Cropper M.,  Vestrand W.~T.,   Ho C.,  2006, \mn@doi
  [The Astrophysical Journal] {10.1086/497327}, 639, 377

\bibitem[\protect\citeauthoryear{Mukhopadhyay \&
  Schaffner-Bielich}{Mukhopadhyay \&
  Schaffner-Bielich}{2016}]{PhysRevD.93.083009}
Mukhopadhyay P.,  Schaffner-Bielich J.,  2016, \mn@doi [Phys. Rev. D]
  {10.1103/PhysRevD.93.083009}, 93, 083009

\bibitem[\protect\citeauthoryear{{Nomoto} \& {Tsuruta}}{{Nomoto} \&
  {Tsuruta}}{1986}]{1986ApJ...305L..19N}
{Nomoto} K.,  {Tsuruta} S.,  1986, \mn@doi [\apjl] {10.1086/184676}, \href
  {https://ui.adsabs.harvard.edu/abs/1986ApJ...305L..19N} {305, L19}

\bibitem[\protect\citeauthoryear{{Nomoto} \& {Tsuruta}}{{Nomoto} \&
  {Tsuruta}}{1987}]{1987ApJ...312..711N}
{Nomoto} K.,  {Tsuruta} S.,  1987, \mn@doi [\apj] {10.1086/164914}, \href
  {https://ui.adsabs.harvard.edu/abs/1987ApJ...312..711N} {312, 711}

\bibitem[\protect\citeauthoryear{Oertel, Hempel, Kl\"ahn  \& Typel}{Oertel
  et~al.}{2017}]{RevModPhys.89.015007}
Oertel M.,  Hempel M.,  Kl\"ahn T.,   Typel S.,  2017, \mn@doi [Rev. Mod.
  Phys.] {10.1103/RevModPhys.89.015007}, 89, 015007

\bibitem[\protect\citeauthoryear{{Ohnishi}, {Tsubakihara}  \&
  {Harada}}{{Ohnishi} et~al.}{2017}]{2017nuco.confb0811O}
{Ohnishi} A.,  {Tsubakihara} K.,   {Harada} T.,  2017, in {Kubono} S.,
  {Kajino} T.,  {Nishimura} S.,  {Isobe} T.,  {Nagataki} S.,  {Shima} T.,
  {Takeda} Y.,  eds, 14th International Symposium on Nuclei in the Cosmos
  (NIC2016). p. 020811, \mn@doi{10.7566/JPSCP.14.020811}

\bibitem[\protect\citeauthoryear{Oppenheimer \& Volkoff}{Oppenheimer \&
  Volkoff}{1939}]{PhysRev.55.374}
Oppenheimer J.~R.,  Volkoff G.~M.,  1939, \mn@doi [Phys. Rev.]
  {10.1103/PhysRev.55.374}, 55, 374

\bibitem[\protect\citeauthoryear{{Page}}{{Page}}{1998}]{1998ASIC..515..539P}
{Page} D.,  1998, in {Buccheri} R.,  {van Paradijs} J.,   {Alpar} A.,  eds,
  NATO Advanced Study Institute (ASI) Series C Vol. 515, The Many Faces of
  Neutron Stars.. p.~539 (\mn@eprint {arXiv} {astro-ph/9706259})

\bibitem[\protect\citeauthoryear{Page \& Reddy}{Page \&
  Reddy}{2006}]{doi:10.1146/annurev.nucl.56.080805.140600}
Page D.,  Reddy S.,  2006, \mn@doi [Annual Review of Nuclear and Particle
  Science] {10.1146/annurev.nucl.56.080805.140600}, 56, 327

\bibitem[\protect\citeauthoryear{{Page}, {Geppert}  \& {Zannias}}{{Page}
  et~al.}{2000}]{2000A&A...360.1052P}
{Page} D.,  {Geppert} U.,   {Zannias} T.,  2000, \aap, \href
  {https://ui.adsabs.harvard.edu/abs/2000A&A...360.1052P} {360, 1052}

\bibitem[\protect\citeauthoryear{Page, Lattimer, Prakash  \& Steiner}{Page
  et~al.}{2004}]{Page_2004}
Page D.,  Lattimer J.~M.,  Prakash M.,   Steiner A.~W.,  2004, \mn@doi [The
  Astrophysical Journal Supplement Series] {10.1086/424844}, 155, 623

\bibitem[\protect\citeauthoryear{Page, Geppert  \& Weber}{Page
  et~al.}{2006}]{PAGE2006497}
Page D.,  Geppert U.,   Weber F.,  2006, \mn@doi [Nuclear Physics A]
  {https://doi.org/10.1016/j.nuclphysa.2005.09.019}, 777, 497

\bibitem[\protect\citeauthoryear{Page, Prakash, Lattimer  \& Steiner}{Page
  et~al.}{2011}]{PhysRevLett.106.081101}
Page D.,  Prakash M.,  Lattimer J.~M.,   Steiner A.~W.,  2011, \mn@doi [Phys.
  Rev. Lett.] {10.1103/PhysRevLett.106.081101}, 106, 081101

\bibitem[\protect\citeauthoryear{Panotopoulos \& Lopes}{Panotopoulos \&
  Lopes}{2017}]{PhysRevD.96.083004}
Panotopoulos G.,  Lopes I.,  2017, \mn@doi [Phys. Rev. D]
  {10.1103/PhysRevD.96.083004}, 96, 083004

\bibitem[\protect\citeauthoryear{Pavlov, Zavlin, Sanwal, Burwitz  \&
  Garmire}{Pavlov et~al.}{2001}]{Pavlov_2001}
Pavlov G.~G.,  Zavlin V.~E.,  Sanwal D.,  Burwitz V.,   Garmire G.~P.,  2001,
  \mn@doi [The Astrophysical Journal] {10.1086/320342}, 552, L129

\bibitem[\protect\citeauthoryear{Pavlov, Zavlin, Sanwal  \& Trümper}{Pavlov
  et~al.}{2002}]{Pavlov_2002}
Pavlov G.~G.,  Zavlin V.~E.,  Sanwal D.,   Trümper J.,  2002, \mn@doi [The
  Astrophysical Journal] {10.1086/340640}, 569, L95

\bibitem[\protect\citeauthoryear{Pethick \& Ravenhall}{Pethick \&
  Ravenhall}{1992}]{10.2307/53909}
Pethick C.~J.,  Ravenhall D.~G.,  1992, Philosophical Transactions: Physical
  Sciences and Engineering, 341, 17

\bibitem[\protect\citeauthoryear{Pons, Reddy, Prakash, Lattimer  \&
  Miralles}{Pons et~al.}{1999}]{Pons_1999}
Pons J.~A.,  Reddy S.,  Prakash M.,  Lattimer J.~M.,   Miralles J.~A.,  1999,
  \mn@doi [\apj] {10.1086/306889}, 513, 780

\bibitem[\protect\citeauthoryear{{Potekhin} \& {Chabrier}}{{Potekhin} \&
  {Chabrier}}{2010}]{2010CoPP...50...82P}
{Potekhin} A.~Y.,  {Chabrier} G.,  2010, \mn@doi [Contributions to Plasma
  Physics] {10.1002/ctpp.201010017}, \href
  {https://ui.adsabs.harvard.edu/abs/2010CoPP...50...82P} {50, 82}

\bibitem[\protect\citeauthoryear{{Potekhin} \& {Chabrier}}{{Potekhin} \&
  {Chabrier}}{2018}]{2018A&A...609A..74P}
{Potekhin} A.~Y.,  {Chabrier} G.,  2018, \mn@doi [\aap]
  {10.1051/0004-6361/201731866}, \href
  {https://ui.adsabs.harvard.edu/abs/2018A&A...609A..74P} {609, A74}

\bibitem[\protect\citeauthoryear{{Potekhin} \& {Yakovlev}}{{Potekhin} \&
  {Yakovlev}}{2001}]{2001A&A...374..213P}
{Potekhin} A.~Y.,  {Yakovlev} D.~G.,  2001, \mn@doi [\aap]
  {10.1051/0004-6361:20010698}, \href
  {https://ui.adsabs.harvard.edu/abs/2001A&A...374..213P} {374, 213}

\bibitem[\protect\citeauthoryear{{Potekhin}, {Chabrier}  \&
  {Yakovlev}}{{Potekhin} et~al.}{1997}]{1997A&A...323..415P}
{Potekhin} A.~Y.,  {Chabrier} G.,   {Yakovlev} D.~G.,  1997, \aap, \href
  {https://ui.adsabs.harvard.edu/abs/1997A&A...323..415P} {323, 415}

\bibitem[\protect\citeauthoryear{{Potekhin}, {Baiko}, {Haensel}  \&
  {Yakovlev}}{{Potekhin} et~al.}{1999}]{Potekhin1999TransportPO}
{Potekhin} A.~Y.,  {Baiko} D.~A.,  {Haensel} P.,   {Yakovlev} D.~G.,  1999,
  \aap, \href {https://ui.adsabs.harvard.edu/abs/1999A&A...346..345P} {346,
  345}

\bibitem[\protect\citeauthoryear{{Potekhin}, {Fantina}, {Chamel}, {Pearson}  \&
  {Goriely}}{{Potekhin} et~al.}{2013}]{2013A&A...560A..48P}
{Potekhin} A.~Y.,  {Fantina} A.~F.,  {Chamel} N.,  {Pearson} J.~M.,   {Goriely}
  S.,  2013, \mn@doi [\aap] {10.1051/0004-6361/201321697}, \href
  {https://ui.adsabs.harvard.edu/abs/2013A&A...560A..48P} {560, A48}

\bibitem[\protect\citeauthoryear{Potekhin, Pons  \& Page}{Potekhin
  et~al.}{2015a}]{Potekhin2015}
Potekhin A.~Y.,  Pons J.~A.,   Page D.,  2015a, \mn@doi [Space Science Reviews]
  {10.1007/s11214-015-0180-9}, 191, 239

\bibitem[\protect\citeauthoryear{{Potekhin}, {Pons}  \& {Page}}{{Potekhin}
  et~al.}{2015b}]{2015SSRv..191..239P}
{Potekhin} A.~Y.,  {Pons} J.~A.,   {Page} D.,  2015b, \mn@doi [\ssr]
  {10.1007/s11214-015-0180-9}, \href
  {https://ui.adsabs.harvard.edu/abs/2015SSRv..191..239P} {191, 239}

\bibitem[\protect\citeauthoryear{{Potekhin}, {Chugunov}  \&
  {Chabrier}}{{Potekhin} et~al.}{2019}]{2019A&A...629A..88P}
{Potekhin} A.~Y.,  {Chugunov} A.~I.,   {Chabrier} G.,  2019, \mn@doi [\aap]
  {10.1051/0004-6361/201936003}, \href
  {https://ui.adsabs.harvard.edu/abs/2019A&A...629A..88P} {629, A88}

\bibitem[\protect\citeauthoryear{Potekhin, Zyuzin, Yakovlev, Beznogov  \&
  Shibanov}{Potekhin et~al.}{2020}]{10.1093/mnras/staa1871}
Potekhin A.~Y.,  Zyuzin D.~A.,  Yakovlev D.~G.,  Beznogov M.~V.,   Shibanov
  Y.~A.,  2020, \mn@doi [Monthly Notices of the Royal Astronomical Society]
  {10.1093/mnras/staa1871}, 496, 5052

\bibitem[\protect\citeauthoryear{Raj, Tanedo  \& Yu}{Raj
  et~al.}{2018}]{PhysRevD.97.043006}
Raj N.,  Tanedo P.,   Yu H.-B.,  2018, \mn@doi [Phys. Rev. D]
  {10.1103/PhysRevD.97.043006}, 97, 043006

\bibitem[\protect\citeauthoryear{{Richardson}, {Savedoff}  \& {van
  Horn}}{{Richardson} et~al.}{1979}]{1979ApJS...39...29R}
{Richardson} M.~B.,  {Savedoff} M.~P.,   {van Horn} H.~M.,  1979, \mn@doi
  [\apjs] {10.1086/190563}, \href
  {https://ui.adsabs.harvard.edu/abs/1979ApJS...39...29R} {39, 29}

\bibitem[\protect\citeauthoryear{{Richardson}, {van Horn}, {Ratcliff}  \&
  {Malone}}{{Richardson} et~al.}{1982}]{1982ApJ...255..624R}
{Richardson} M.~B.,  {van Horn} H.~M.,  {Ratcliff} K.~F.,   {Malone} R.~C.,
  1982, \mn@doi [\apj] {10.1086/159865}, \href
  {https://ui.adsabs.harvard.edu/abs/1982ApJ...255..624R} {255, 624}

\bibitem[\protect\citeauthoryear{Ring, Gambhir  \& Lalazissis}{Ring
  et~al.}{1997}]{RING199777}
Ring P.,  Gambhir Y.,   Lalazissis G.,  1997, \mn@doi [Computer Physics
  Communications] {https://doi.org/10.1016/S0010-4655(97)00022-2}, 105, 77

\bibitem[\protect\citeauthoryear{Ruffert \& Janka}{Ruffert \&
  Janka}{2010}]{refId0}
Ruffert M.,  Janka H.~T.,  2010, \mn@doi [A\&A] {10.1051/0004-6361/200912738},
  514, A66

\bibitem[\protect\citeauthoryear{Safi-Harb \& Kumar}{Safi-Harb \&
  Kumar}{2008}]{Safi_Harb_2008}
Safi-Harb S.,  Kumar H.~S.,  2008, \mn@doi [The Astrophysical Journal]
  {10.1086/590359}, 684, 532

\bibitem[\protect\citeauthoryear{Serot \& Walecka}{Serot \&
  Walecka}{1992}]{Serot1992}
Serot B.~D.,  Walecka J.~D.,  1992, Relativistic Nuclear Many-Body Theory.
Springer US, Boston, MA, pp 49--92, \mn@doi{10.1007/978-1-4615-3466-2_5}, \url
  {https://doi.org/10.1007/978-1-4615-3466-2_5}

\bibitem[\protect\citeauthoryear{Shen}{Shen}{2002}]{PhysRevC.65.035802}
Shen H.,  2002, \mn@doi [Phys. Rev. C] {10.1103/PhysRevC.65.035802}, 65, 035802

\bibitem[\protect\citeauthoryear{Shternin \& Yakovlev}{Shternin \&
  Yakovlev}{2008}]{PhysRevD.78.063006}
Shternin P.~S.,  Yakovlev D.~G.,  2008, \mn@doi [Phys. Rev. D]
  {10.1103/PhysRevD.78.063006}, 78, 063006

\bibitem[\protect\citeauthoryear{Shternin, Baldo  \& Haensel}{Shternin
  et~al.}{2013}]{PhysRevC.88.065803}
Shternin P.~S.,  Baldo M.,   Haensel P.,  2013, \mn@doi [Phys. Rev. C]
  {10.1103/PhysRevC.88.065803}, 88, 065803

\bibitem[\protect\citeauthoryear{Slattery, Doolen  \& DeWitt}{Slattery
  et~al.}{1982}]{PhysRevA.26.2255}
Slattery W.~L.,  Doolen G.~D.,   DeWitt H.~E.,  1982, \mn@doi [Phys. Rev. A]
  {10.1103/PhysRevA.26.2255}, 26, 2255

\bibitem[\protect\citeauthoryear{{Thorne}}{{Thorne}}{1977}]{1977ApJ...212..825T}
{Thorne} K.~S.,  1977, \mn@doi [\apj] {10.1086/155108}, \href
  {https://ui.adsabs.harvard.edu/abs/1977ApJ...212..825T} {212, 825}

\bibitem[\protect\citeauthoryear{Todd-Rutel \& Piekarewicz}{Todd-Rutel \&
  Piekarewicz}{2005}]{PhysRevLett.95.122501}
Todd-Rutel B.~G.,  Piekarewicz J.,  2005, \mn@doi [Phys. Rev. Lett.]
  {10.1103/PhysRevLett.95.122501}, 95, 122501

\bibitem[\protect\citeauthoryear{Tolman}{Tolman}{1939}]{PhysRev.55.364}
Tolman R.~C.,  1939, \mn@doi [Phys. Rev.] {10.1103/PhysRev.55.364}, 55, 364

\bibitem[\protect\citeauthoryear{{Urpin}, {Konenkov}  \& {Urpin}}{{Urpin}
  et~al.}{1997}]{1997MNRAS.292..167U}
{Urpin} V.,  {Konenkov} D.,   {Urpin} V.,  1997, \mn@doi [\mnras]
  {10.1093/mnras/292.1.167}, \href
  {https://ui.adsabs.harvard.edu/abs/1997MNRAS.292..167U} {292, 167}

\bibitem[\protect\citeauthoryear{Waldhauser, Theis, Maruhn, St\"ocker  \&
  Greiner}{Waldhauser et~al.}{1987}]{PhysRevC.36.1019}
Waldhauser B.~M.,  Theis J.,  Maruhn J.~A.,  St\"ocker H.,   Greiner W.,  1987,
  \mn@doi [Phys. Rev. C] {10.1103/PhysRevC.36.1019}, 36, 1019

\bibitem[\protect\citeauthoryear{Walecka}{Walecka}{1974}]{WALECKA1974491}
Walecka J.,  1974, \mn@doi [Annals of Physics]
  {https://doi.org/10.1016/0003-4916(74)90208-5}, 83, 491

\bibitem[\protect\citeauthoryear{Weissenborn, Chatterjee  \&
  Schaffner-Bielich}{Weissenborn et~al.}{2012}]{PhysRevC.85.065802}
Weissenborn S.,  Chatterjee D.,   Schaffner-Bielich J.,  2012, \mn@doi [Phys.
  Rev. C] {10.1103/PhysRevC.85.065802}, 85, 065802

\bibitem[\protect\citeauthoryear{Xiang, Jiang, Zhang  \& Yang}{Xiang
  et~al.}{2014}]{PhysRevC.89.025803}
Xiang Q.-F.,  Jiang W.-Z.,  Zhang D.-R.,   Yang R.-Y.,  2014, \mn@doi [Phys.
  Rev. C] {10.1103/PhysRevC.89.025803}, 89, 025803

\bibitem[\protect\citeauthoryear{{Yakovlev} \& {Levenfish}}{{Yakovlev} \&
  {Levenfish}}{1995}]{1995A&A...297..717Y}
{Yakovlev} D.~G.,  {Levenfish} K.~P.,  1995, \aap, \href
  {https://ui.adsabs.harvard.edu/abs/1995A&A...297..717Y} {297, 717}

\bibitem[\protect\citeauthoryear{{Yakovlev} \& {Urpin}}{{Yakovlev} \&
  {Urpin}}{1980}]{1980SvA....24..303Y}
{Yakovlev} D.~G.,  {Urpin} V.~A.,  1980, \sovast, \href
  {https://ui.adsabs.harvard.edu/abs/1980SvA....24..303Y} {24, 303}

\bibitem[\protect\citeauthoryear{Yakovlev, Levenfish  \& Shibanov}{Yakovlev
  et~al.}{1999}]{Yakovlev_1999}
Yakovlev D.~G.,  Levenfish K.~P.,   Shibanov Y.~A.,  1999, \mn@doi
  [Physics-Uspekhi] {10.1070/pu1999v042n08abeh000556}, 42, 737–778

\bibitem[\protect\citeauthoryear{Yakovlev, Kaminker, Gnedin  \&
  Haensel}{Yakovlev et~al.}{2001}]{YAKOVLEV20011}
Yakovlev D.,  Kaminker A.,  Gnedin O.,   Haensel P.,  2001, \mn@doi [Physics
  Reports] {https://doi.org/10.1016/S0370-1573(00)00131-9}, 354, 1

\bibitem[\protect\citeauthoryear{Zavlin}{Zavlin}{2007a}]{Zavlinbook_2007}
Zavlin V.~E.,  2007a, in {Neutron Stars and Pulsars: About 40 Years After the
  Discovery: 363rd Heraeus Seminar}.  (\mn@eprint {arXiv} {astro-ph/0702426})

\bibitem[\protect\citeauthoryear{Zavlin}{Zavlin}{2007b}]{Zavlin_2007}
Zavlin V.~E.,  2007b, \mn@doi [The Astrophysical Journal] {10.1086/521300},
  665, L143

\bibitem[\protect\citeauthoryear{Zavlin \& Pavlov}{Zavlin \&
  Pavlov}{2004}]{Zavlin_2004}
Zavlin V.~E.,  Pavlov G.~G.,  2004, Mem. Soc. Ast. It., 75, 458

\bibitem[\protect\citeauthoryear{Zavlin, Trumper  \& Pavlov}{Zavlin
  et~al.}{1999}]{Zavlin_1999}
Zavlin V.~E.,  Trumper J.,   Pavlov G.~G.,  1999, \mn@doi [The Astrophysical
  Journal] {10.1086/307919}, 525, 959

\bibitem[\protect\citeauthoryear{Zhang \& Chen}{Zhang \&
  Chen}{2016}]{PhysRevC.94.064326}
Zhang Z.,  Chen L.-W.,  2016, \mn@doi [Phys. Rev. C]
  {10.1103/PhysRevC.94.064326}, 94, 064326

\bibitem[\protect\citeauthoryear{de Lavallaz \& Fairbairn}{de~Lavallaz \&
  Fairbairn}{2010}]{PhysRevD.81.123521}
de Lavallaz A.,  Fairbairn M.,  2010, \mn@doi [Phys. Rev. D]
  {10.1103/PhysRevD.81.123521}, 81, 123521

\bibitem[\protect\citeauthoryear{{van Riper}}{{van
  Riper}}{1991}]{1991ApJS...75..449V}
{van Riper} K.~A.,  1991, \mn@doi [\apjs] {10.1086/191538}, \href
  {https://ui.adsabs.harvard.edu/abs/1991ApJS...75..449V} {75, 449}

\bibitem[\protect\citeauthoryear{{van Riper} \& {Lamb}}{{van Riper} \&
  {Lamb}}{1981}]{1981ApJ...244L..13V}
{van Riper} K.~A.,  {Lamb} D.~Q.,  1981, \mn@doi [\apjl] {10.1086/183469},
  \href {https://ui.adsabs.harvard.edu/abs/1981ApJ...244L..13V} {244, L13}

\makeatother
\end{thebibliography}
\bibliographystyle{mnras}
\end{document}